\documentclass[manuscript,screen,bookmarksnumbered,unicode,hidelinks]{acmart}

\usepackage{amsmath}
\usepackage{array}
\usepackage{graphicx}
\usepackage{tabularx}
\usepackage{makecell}
\usepackage{fontawesome5}
\usepackage{tcolorbox}
\usepackage{enumitem}
\tcbuselibrary{skins,breakable}
\newcolumntype{Y}{>{\centering\arraybackslash}X}
\usepackage{placeins}
\usepackage{float}
\usepackage{multirow}
\usepackage{dblfloatfix}
\usepackage{caption}
\usepackage{comment}
\usepackage{needspace} 
\setlist{nosep}

\renewcommand\footnotetextcopyrightpermission[1]{}
\renewcommand\footnotetextcopyrightpermission[1]{}
\newcommand{\myfootertext}{Accepted in ACM Transactions on Computing Education (TOCE)}
\makeatletter
\AtBeginDocument{%
  \let\ACMorigfirst\ps@firstpagestyle
  \def\ps@firstpagestyle{\ACMorigfirst
    \fancyfoot[RE,LO]{\footnotesize\myfootertext}}%
  \let\ACMorigstd\ps@standardpagestyle
  \def\ps@standardpagestyle{\ACMorigstd
    \fancyfoot[RO,LE]{\footnotesize\myfootertext}}%
  \pagestyle{standardpagestyle}%
}
\makeatother

\title{Motivators and Demotivators of Integrating Large Language Models in Software Engineering Education: An Exploratory Familiarity Allocation Framework}

\author{Maryam Khan}
\affiliation{%
  \institution{Department of Software Engineering, LUT University}
  \city{Lappeenranta}
  \country{Finland}}
\email{maryam.khan@student.lut.fi}

\author{Jussi Kasurinen}
\affiliation{%
  \institution{Department of Software Engineering, LUT University}
  \city{Lappeenranta}
  \country{Finland}}
\email{jussi.kasurinen@lut.fi}

\author{Muhammad Azeem Akbar}
\affiliation{%
  \institution{Department of Software Engineering, LUT University}
  \city{Lappeenranta}
  \country{Finland}}
\email{azeem.akbar@lut.fi}
\author{Estefanía Martín-Barroso}
\affiliation{%
  \institution{Escuela Técnica Superior de Ingeniería Informática, Universidad Rey Juan Carlos}
  \city{Móstoles}
  \country{Spain}}
\email{estefania.martin@urjc.es}

\begin{document}

\begin{abstract}
Large Language Models (LLMs) are increasingly influencing software engineering practice and education. However, limited research has examined their systematic integration using stakeholder evidence and quantitative decision-support models. This study presents an exploratory familiarity--allocation framework for analyzing motivating and demotivating factors associated with LLM integration in software engineering education. Building on our previously developed literature-based taxonomy, we evaluated 19 factors, comprising nine motivators and ten demotivators, through a structured survey of 126 stakeholders from multiple countries. Naive Bayes and Logistic Regression were used to examine associations between the factor ratings and high self-reported LLM familiarity. Both models achieved an accuracy of 0.461, below the majority-class baseline, indicating limited ability to distinguish respondents with higher and lower familiarity. Their estimated probabilities were therefore used as exploratory relative scoring functions rather than reliable individual-level predictions. These estimates were incorporated into a Genetic Algorithm using a defined allocation score to support Global Factor Allocation (GC), Local Factor Allocation (LC), and category-level $\Delta$Fitness analyses. The descriptive results identified \textit{Programming Assistance and Debugging Support}, \textit{Personalized and Adaptive Learning}, and \textit{AI as a Learning Partner} as the most highly rated motivators. The most highly rated demotivators were \textit{Plagiarism and Intellectual Property Concerns}, \textit{Over-Reliance on AI in Learning}, and \textit{Reduced Critical Thinking and Problem-Solving}. Under the fitted probability models and allocation assumptions, GC assigned relatively higher levels to integrity- and governance-related factors, while LC revealed additional pedagogical and risk-related patterns within individual categories. The resulting configurations represent exploratory familiarity--allocation scenarios rather than validated implementation priorities, institutional costs, or evidence of educational effectiveness. Classroom and institutional validation is required before the framework can guide LLM integration in practice.

\end{abstract}


\maketitle

\section{Introduction}

Artificial Intelligence (AI) is increasingly influencing 
education. Zhai et al.~\cite{zhai2021review} reviewed AI applications in 
education, while Jin et al.~\cite{jin2025generative} examined institutional 
policies and guidelines for adopting generative AI. Recent advances in natural 
language processing and deep learning have led to the development of Large 
Language Models (LLMs), which can generate coherent and contextually relevant 
responses \cite{zhao2023survey}. Although many contemporary LLMs rely on 
Transformer-based architectures and large-scale pre-training, their development 
does not follow a single architecture or training process. Models vary in their 
architectural configurations, training data, pre-training objectives, scale, 
and adaptation and post-training methods, including fine-tuning, instruction 
tuning, and alignment using human feedback 
\cite{zhao2023survey,bommasani2021opportunities}. These choices collectively 
influence their capabilities and limitations. Their use also raises ethical 
concerns that require careful consideration \cite{akbar2023ethical}. This study 
focuses specifically on the integration of LLMs into software engineering 
education.

The rapid adoption of LLMs creates both important opportunities and significant challenges for software engineering education \cite{he2025llm}. LLM-based tools are already integrated into professional software development, supporting activities such as system design, code implementation, testing, and project management \cite{kharrufa2026llms}. As industry practices evolve toward AI-augmented development workflows, software engineering education must adapt accordingly. Educational programs therefore require systematic revision to ensure effective, responsible, and pedagogically sound integration of LLM technologies \cite{fan2023large, kirova2024software}. Universities must adopt structured strategies that protect academic integrity while helping students and educators understand the benefits, limitations, and risks associated with LLM use \cite{daun2023chatgpt}.

Several studies have explored the adoption of AI technologies in computer science and software engineering education \cite{banerjee2025understanding, feldt2018ways, lo2023impact}. However, the rapidly evolving AI landscape introduces new pedagogical and curricular challenges that extend beyond initial adoption. Defining appropriate institutional strategies, assessment models, and learning outcomes in AI-augmented environments remains complex. The systematic and responsible integration of advanced AI technologies, such as LLMs, into software engineering education is still underexplored and continues to be an active area of research \cite{bommasani2021opportunities, khan2025integrating, becker2023programming}.

Educational systems that promote innovation and progressive teaching practices provide a valuable setting for integrating AI tools and establishing structured usage guidelines \cite{fattahibavandpour2024advancing}. Integration efforts should align with institutional goals, instructional practices, and regulatory frameworks. Recent research emphasizes the importance of human-centered AI approaches in education, particularly in supporting educators through AI-enabled course design and management \cite{macias2024empowering}. Such approaches help instructors manage complex learning environments while encouraging responsible and context-aware use of LLMs.

At the same time, LLM integration raises important concerns. Issues such as biased outputs, ethical implications, and the opaque “black-box” nature of these models have been widely discussed \cite{akbar2023ethical, bender2021dangers}. These concerns extend beyond academic integrity and have practical implications for software engineering students who increasingly interact with and develop AI-driven systems. Ethical principles such as fairness, privacy, accountability, and transparency are therefore critical considerations \cite{khan2023ai}. Embedding these principles into software engineering education is essential, as future professionals will design and deploy AI-enabled systems that affect society \cite{akbar2023ethical}.

Despite global digital transformation efforts in higher education, LLM adoption remains uneven and often lacks structured implementation strategies. Although AI tools are increasingly used for course planning and content development, their systematic integration into software engineering curricula is still limited and insufficiently supported by empirical evidence. This gap highlights the need for rigorous, evidence-based research to guide responsible and effective LLM integration. Integrating LLM-related considerations into software engineering curricula is therefore essential \cite{khan2022ethics}. Addressing this challenge requires a comprehensive understanding of both pedagogical implications and the technical characteristics of LLM technologies.

To establish the conceptual foundation for the present study,
our previously published literature review \cite{khan2025integrating}
systematically synthesized evidence on LLM integration in software engineering
education and developed taxonomies of motivating and demotivating factors. The
present study builds on these taxonomies by operationalizing their sub-themes
as 19 survey factors and empirically evaluating them through a
multi-stakeholder survey. Based on the survey data, we developed an exploratory
probabilistic and allocation framework to examine associations between
stakeholder perceptions and high self-reported LLM familiarity. The estimated
probabilities were subsequently combined with a defined allocation score to
compare alternative factor configurations. The optimization component follows
the established Genetic Algorithm principles described by Goldberg
\cite{Goldberg1989}.

The modeled outcome represents familiarity rather than successful curriculum integration. Prediction and optimization models have previously been applied in software engineering to support decision-making in complex development environments. For example, prior studies have employed machine learning algorithms such as Support Vector Machines, Artificial Neural Networks, and Random Forests to develop predictive frameworks for identifying key DevOps practices and estimating the likelihood of successful DevOps implementation in software organizations \cite{kumar2023machine}. Similarly, metaheuristic-based optimization techniques combined with probabilistic classifiers have been used to develop cost-effective predictive models for DevOps project success by balancing success probability and implementation cost \cite{kumar2024metaheuristic}. In emerging domains, optimization-based models have also been proposed to predict the success of quantum software development projects, demonstrating how evolutionary algorithms can improve success probability while accounting for project cost and complexity \cite{khan2024agile}. However, based on our understanding of the state-of-the-art literature 
(Section~\ref{sec:RelatedWork}), such probabilistic and optimization-driven 
approaches have not yet been applied to guide the systematic integration of 
emerging AI technologies, such as LLMs, within software engineering education. 

The overall objective of this study is to empirically evaluate the
literature-derived taxonomy of motivating and demotivating factors across
multiple stakeholder groups and to develop an exploratory
familiarity--allocation framework. Specifically, the study examines the
associations between the 19 factor ratings and high self-reported LLM
familiarity and uses the estimated probabilities with a defined allocation
score to compare alternative factor configurations. This exploratory study
examines associations rather than testing a predefined causal relationship;
therefore, no formal hypothesis is proposed. Accordingly, the study is guided
by the following research question: 
\begin{itemize}
    \item \textit{\textbf{RQ:} How can configurations of motivating and demotivating factors be compared across global and category levels under an exploratory familiarity--allocation objective in software engineering education?}
\end{itemize}

By integrating stakeholder perceptions, probabilistic modeling, and
evolutionary optimization, the proposed framework provides an exploratory
mechanism for comparing GC, LC and category-level $\Delta$Fitness under the fitted probability models
and the defined allocation score. The framework offers preliminary
decision-support insights rather than reliable individual-level predictions
or validated implementation guidance.

The remainder of this paper is structured as follows.
Section~\ref{sec:RelatedWork} reviews related work and positions the study
against existing research. Section~\ref{sec:Methodology} describes the research
methodology, including the literature-derived taxonomy, survey design, data
preparation, probabilistic modeling, and GA-based allocation framework.
Section~\ref{Sec4:Results} presents the descriptive, probabilistic, and
allocation results. Section~\ref{sec:Discussions} discusses the findings and
interprets the exploratory framework. Section~\ref{sec:implications} presents
the academic, instructional, and institutional implications.
Section~\ref{sec:Threats} discusses threats to validity. Finally,
Section~\ref{sec:conclusion} concludes the paper and outlines directions for
future research.
\section{Related Work and Motivation} \label{sec:RelatedWork}
The existing literature presents a growing body of work on the integration of LLMs in education broadly, and in software engineering education in particular. To systematically synthesize the current state of the art and identify the research gap that motivates this study, we categorize prior research into three main streams: (1) LLM-based tools evaluation, (2) classroom and student-centered empirical investigations, and (3) curriculum adaptation and pedagogical transformation. The details of each category are discussed below.

\subsection{LLM-based Tools Evaluation} \label{Sec:2.1}
This stream of research studies focuses on evaluating the technical capabilities, reliability, and instructional effectiveness of LLM-based tools in software engineering education. Within this context, several studies have conducted systematic performance assessments of different LLM platforms to determine their suitability for educational use. For instance, Pereira et al.\cite{pereira2024leveraging}  conducted a comparative evaluation of generative AI platforms, specifically ChatGPT, Mistral, and Llama, within the context of software engineering education. Using a set of representative tasks derived from the Software Engineering Body of Knowledge (SWEBOK) framework \cite{washizaki2024guide}, they examined the performance of these models across various software engineering activities. The findings indicate that ChatGPT and Mistral generally outperform Llama in terms of response quality and task execution; however, none of the models demonstrate consistent reliability when handling complex or cognitively demanding tasks. The study concludes that while LLM-based tools show promise for supporting software engineering education, instructors must exercise caution and systematically verify model-generated outputs before classroom adoption.

Similarly, efforts to enhance reliability and instructional quality through architectural improvements are discussed in the study conducted by Song et al.\cite{song2024high} which proposes a Customized Role-Based Agent (CRBA) framework designed to enhance the quality of programming projects generated using LLM-based tools for educational purposes. The framework employs an iterative multi-agent architecture in which specialized agents critique and refine intermediate outputs to address reliability concerns associated with complex task generation. The approach was evaluated using expert-defined quality metrics and a questionnaire survey involving 60 undergraduate students. The results indicate that LLM-generated projects perform comparably to instructor-designed projects in terms of clarity, learning structure, and overall quality. The findings demonstrate the potential of structured, role-based LLM systems to improve the scalability and efficiency of developing high-quality programming assignments for educational settings.

Beyond content and project generation, LLM-based systems have also been integrated into learning platforms to support student interaction and engagement. Neumann et al.\cite{neumann2024llm} developed an LLM-based chatbot, MoodleBot, which was integrated into a university learning management system to support self-regulated learning in a database course. The system was built using GPT-4 and a retrieval-augmented generation (RAG) architecture, and its effectiveness was evaluated through the Technology Acceptance Model (TAM). The results indicate high levels of student acceptance and an overall response accuracy of approximately 88\%, although limitations were observed in detecting incorrect or misleading responses. The study highlights the pedagogical potential of LLM-based chatbots while also addressing their operational and reliability challenges within higher education environments.

In parallel, benchmarking studies have assessed the raw performance of large language models on programming tasks. For instance, Finnie-Ansley et al.\cite{finnie2022robots} investigate the performance of OpenAI Codex on introductory programming tasks. Using real CS1 exam questions and multiple variants of the well-known Rainfall problem, they compare Codex-generated solutions against student performance. Their results show that Codex solves the majority of exam-style problems successfully and would rank within the top quartile of students under standard grading conditions. However, the model struggles with tightly constrained tasks and specific formatting requirements, while producing substantial variation in solution structure. The authors discuss significant implications for assessment design, academic integrity, and the future of computing education in the presence of AI code generation tools.

\subsection{Classroom and Student-Centered Studies} \label{Sec:2.2}
In this category of related work, we encapsulated studies that examine the real-world deployment of LLM-based tools with empirical evidence from classroom environments, evaluating both performance improvements and potential risks associated with LLM adoption. For instance, Lyu et al. \cite{lyu2024evaluating} conducted a semester-long study between students to examine the impact of an LLM-powered assistant, CodeTutor, in an entry-level programming course. Students who used CodeTutor achieved statistically significant improvements in final scores compared to the control group, with greater gains observed among those without prior experience with LLM-based tools. Participants positively evaluated the tool effectiveness in supporting syntax learning and debugging tasks. However, concerns were identified regarding its limited contribution to the development of critical thinking skills, and students reliance on the tool over time. The study underscores the importance of AI literacy, prompt quality, and sustained pedagogical strategies when integrating LLM-powered assistants into programming education.

Extending the investigation to collaborative learning contexts, the study conducted by Kharrufa et al.\cite{kharrufa2026llms} explores the role of LLMs in team-based software engineering projects. The authors investigated the use of generative AI tools, including ChatGPT and GitHub Copilot, within a semester-long team-based software engineering project. Data were collected through surveys and interviews to examine the implications of LLM-based tools on students’ productivity, learning experiences, and teamwork dynamics. The findings indicate that generative AI tools can enhance productivity, self-efficacy, and student confidence. However, their use also introduces challenges, such as overreliance on AI, reduced accountability, and the emergence of hidden skill gaps. The authors further propose a pedagogically informed design space to support the structured and transparent integration of generative AI tools into computing education.

Similarly, Kazemitabaar et al.\cite{kazemitabaar2023studying} examine the impact of AI-based code generation tools on novice programmers in an introductory computing course. Through an experimental study, the authors compare students who used an AI coding assistant with those who did not, measuring task performance, learning outcomes, and engagement. The findings indicate that AI support can improve task completion efficiency and short-term performance; however, concerns emerge regarding overreliance and uneven learning gains across students. The study highlights the need for carefully designed instructional strategies to ensure that AI tools enhance, rather than undermine, conceptual understanding in programming education.

Recent studies have also examined LLM use in software 
quality assurance education. Maxim et al. \cite{maxim2025generative} 
introduced ChatGPT-assisted debugging in a hybrid undergraduate video game 
project course. Students used ChatGPT to suggest repairs for defects in a 
Unity 3D scene and reflected on the quality of its recommendations. Most 
students completed the debugging tasks and felt that ChatGPT improved their 
efficiency and understanding of the development environment. AlOmar and 
Mkaouer \cite{alomar2024cultivating} combined ChatGPT with the Programming 
Mistake Detector in a software quality course involving 102 submissions 
and 1,230 code-quality issues. Their findings show that ChatGPT can support 
code review and debugging, but its recommendations may lack context or 
introduce incorrect changes. They therefore emphasize the need for students 
to verify LLM-generated suggestions using human judgment and traditional 
analysis tools. These studies provide practical evidence from specific 
software quality activities, whereas our study examines broader 
institutional and pedagogical factors affecting LLM integration in software 
engineering education.

\subsection{Curriculum Adaptation and Pedagogical Transformation} \label{Sec:2.3}
We also included related research studies focusing on the curricular and pedagogical implications of LLM integration. These studies examine how learning outcomes, assessment strategies, and instructional design must evolve to align with AI-augmented development environments. Such a perspective is presented by Kirova et al.\cite{kirova2024software} who argue that software engineering education must adapt to the growing influence of AI and, more specifically LLM. The paper discusses how LLMs can be integrated into curriculum design, assessment practices, and conceptual instruction, as well as how they may enhance technical activities such as testing, validation, and verification. In addition, the authors examine associated risks, including bias, intellectual property concerns, and hallucinations, and advocate for the explicit incorporation of responsible AI principles into software engineering curricula. They ultimately call for a holistic approach that integrates technical competence, professional responsibility, and critical thinking to prepare students for AI-driven development environments.

Similarly, a study conducted by  Zönnchen et al.\cite{zonnchen2024impact} examines how the rapid emergence of ChatGPT necessitates reconsideration of teaching and assessment strategies in software engineering. The authors argue that the emergence of ChatGPT may render certain traditional skills less central while necessitating the development of new competencies among future software engineers. They emphasize the need to revise intended learning outcomes, assessment strategies, and teaching methods to maintain constructive alignment in AI-augmented educational environments. The paper calls for systematic curricular adaptation to address the evolving skill requirements of software engineering in the era of generative AI.

Expanding this discussion to computing education more broadly, the study conducted by Denny et al.\cite{denny2024computing} discusses the transformative impact of generative AI and code-generation models on computing education. Drawing on empirical evaluations of models such as Codex and GPT-4, the authors show that these systems can successfully solve many introductory programming tasks, raising concerns about academic integrity, overreliance, bias, and secure coding practices. At the same time, the paper highlights emerging opportunities, including automated generation of programming exercises, code explanations, improved error messages, and new pedagogical approaches centered on prompt engineering and specification skills. The authors argue that computing curricula must evolve to integrate AI tools responsibly while preserving core competencies in code literacy, critical evaluation, and software verification.

Raihan et al. \cite{raihan2025large} conducted a 
systematic literature review of LLM use in computer science education. 
After screening 1,735 records, they included 125 studies covering 
educational levels, computer science disciplines, research methods, 
programming languages, and LLMs. Their findings show that existing research 
mainly focuses on undergraduate education and introductory programming. 
Although students generally view LLMs positively, instructors remain 
concerned about over-reliance, plagiarism, and possible negative effects 
on independent learning. The review also highlights the need to adapt 
computing curricula to recent AI developments. Our study complements this 
work by focusing specifically on software engineering education and by 
examining motivators, demotivators, familiarity, and exploratory familiarity–allocation analysis through a multi-stakeholder survey and an exploratory 
optimization framework.

\subsection{Motivation}
This subsection explains the motivation for the present study 
based on the related work discussed in Sections~2.1--2.3. Table~\ref{tab:related_work_comparison} compares previous studies and the
present study across six dimensions: motivators, demotivators, taxonomy
development, empirical evaluation, multi-stakeholder perspective, and
exploratory allocation modeling. These dimensions were selected based on our
previous literature review \cite{khan2025integrating}, the gaps identified in
the related work, and the focus of the present study. Motivators and
demotivators indicate whether previous studies examined the expected benefits
and challenges of LLM integration. Taxonomy development indicates whether
these factors were systematically organized, while empirical evaluation shows
whether they were examined using stakeholder data. The multi-stakeholder
dimension indicates whether different stakeholder groups were included.
Exploratory allocation modeling indicates whether stakeholder perceptions were
combined with a quantitative framework for comparing alternative factor
configurations under explicitly defined modeling assumptions.

\begin{table}[ht]
\centering
\footnotesize
\setlength{\tabcolsep}{1pt}

\caption{Comparison of prior studies and our work across key dimensions.}
\label{tab:related_work_comparison}

\begin{tabular}{|p{4.5cm}|c|c|c|c|c|c|}
\hline
\textbf{Studies} &
\textbf{Motivator} &
\textbf{Demotivator} &
\textbf{\shortstack{Taxonomy\\Development}} &
\textbf{\shortstack{Empirical\\Validation}} &
\textbf{\shortstack{Multi-\\Stakeholders}} &
\textbf{\shortstack{Exploratory--Allocation\\Model}} \\
\hline

Pereira et al.~\cite{pereira2024leveraging} & X & $\checkmark$ (+) & $\checkmark$ (+) & $\checkmark$ & X & X \\
\hline
Song et al.~\cite{song2024high} & X & $\checkmark$ (+) & X & $\checkmark$ & X & X \\
\hline
Neumann et al.~\cite{neumann2024llm} & $\checkmark$ (+) & $\checkmark$ (+) & X & $\checkmark$ & $\checkmark$ & X \\
\hline
Finnie-Ansley et al.~\cite{finnie2022robots} & $\checkmark$ (+) & $\checkmark$ & X & $\checkmark$ & X & X \\
\hline
Lyu et al.~\cite{lyu2024evaluating} & $\checkmark$ (+) & $\checkmark$ (+) & X & $\checkmark$ & X & X \\
\hline
Kharrufa et al.~\cite{kharrufa2026llms} & $\checkmark$ & $\checkmark$ & X & $\checkmark$ & X & X \\
\hline
Kazemitabaar et al.~\cite{kazemitabaar2023studying} & $\checkmark$ (+) & $\checkmark$ (+) & X & $\checkmark$ & X & X \\
\hline
Kirova et al.~\cite{kirova2024software} & $\checkmark$ (+) & $\checkmark$ (+) & X & X & X & X \\
\hline
Zönnchen et al.~\cite{zonnchen2024impact} & $\checkmark$ & $\checkmark$ & $\checkmark$ (+) & $\checkmark$ & $\checkmark$ (+) & X \\
\hline
Denny et al.~\cite{denny2024computing} & $\checkmark$ & $\checkmark$ & X & $\checkmark$ (+) & X & X \\
\hline
Maxim et al.~\cite{maxim2025generative} & $\checkmark$ & $\checkmark$ (+) & X & $\checkmark$ & X & X \\
\hline
AlOmar and Mkaouer~\cite{alomar2024cultivating} & $\checkmark$ & $\checkmark$ & X & $\checkmark$ & X & X \\
\hline
Raihan et al.~\cite{raihan2025large} & $\checkmark$ & $\checkmark$ & X & $\checkmark$ (+) & $\checkmark$ (+) & X \\
\hline
\textbf{Our Study} & $\checkmark$ & $\checkmark$ & $\checkmark$ & $\checkmark$ & $\checkmark$ & $\checkmark$ \\
\hline
\multicolumn{7}{|l|}{\scriptsize \textit{Legend:} X = Not discussed; $\checkmark$ = Discussed; $\checkmark$ (+) = Partially discussed.} \\
\hline

\end{tabular}
\end{table}

First, several prior studies have examined either the technical performance or classroom impact of LLM-based tools. For example, Pereira et al.\cite{pereira2024leveraging} and Finnie-Ansley et al.\cite{finnie2022robots} conducted benchmarking analyses to evaluate model reliability and programming performance. Similarly, Song et al.\cite{song2024high} proposed architectural improvements to enhance the quality of LLM-generated programming tasks. While these works provide valuable empirical validation of tool capabilities, they largely focus on system-level performance evaluation. However, these studies do not systematically model motivating or demotivating factors influencing educational integration. Moreover, none of these studies combines probabilistic familiarity estimates 
with exploratory familiarity–allocation analysis to explore the prioritization of motivating 
and demotivating factors in software engineering education.

Second, classroom-centered empirical investigations such as those by Lyu 
et al.\cite{lyu2024evaluating}, Kharrufa et al.\cite{kharrufa2026llms}, 
and Kazemitabaar et al.\cite{kazemitabaar2023studying} provide meaningful 
insights into student performance, productivity, and learning outcomes. 
The studies by Maxim et al.
\cite{maxim2025generative} and AlOmar and Mkaouer
\cite{alomar2024cultivating} extend this evidence to software quality 
assurance education. They show how LLMs can support debugging, code review, 
and defect correction, while also identifying problems such as limited 
code context and incorrect AI-generated recommendations. However, both 
studies focus on specific classroom activities and student experiences 
rather than broader institutional integration.
These studies discuss both positive impacts (e.g., improved task efficiency, 
engagement) and negative implications (e.g., overreliance, reduced critical 
thinking). However, as shown in Table 
\ref{tab:related_work_comparison}, motivators and demotivators are typically 
discussed descriptively rather than structured into formal taxonomies 
and connected with probabilistic and exploratory optimization 
analyses. Additionally, these investigations primarily involve 
student participants, limiting the perspective to a single stakeholder 
group and restricting generalizability to institutional decision-making 
contexts.

Third, curriculum-oriented discussions, including those by Kirova et 
al.\cite{kirova2024software}, Zönnchen et 
al.\cite{zonnchen2024impact} and Denny et.
al.\cite{denny2024computing} emphasize the need for pedagogical 
transformation in response to AI-augmented development environments. Kirova 
et al.\cite{kirova2024software} provide a conceptual and review-based 
perspective on curricular adaptation, highlighting opportunities and risks 
without empirical validation. In contrast, Zönnchen et 
al.\cite{zonnchen2024impact} offer empirical insights and partially address 
taxonomy-related aspects and stakeholder considerations, while Denny et 
al.\cite{denny2024computing} provide partial empirical validation through 
model performance analyses and synthesized evidence. 
Similarly, Raihan et al.
\cite{raihan2025large} synthesize 125 studies on LLM use in computer 
science education. Their review identifies educational benefits as well as 
concerns related to over-reliance, plagiarism, independent learning, and 
curriculum adaptation. However, it provides a broad review of computing education and does not combine 
software engineering-specific factors with probabilistic familiarity modeling 
and exploratory familiarity–allocation analysis.
Nevertheless, despite these contributions, none of these studies develops 
a comprehensive, empirically grounded taxonomy of motivating and 
demotivating factors combined with probabilistic familiarity modeling 
and exploratory familiarity–allocation analysis in software engineering education.

Bayesian classification has also been applied in computing education and 
software engineering decision-support research. Veerasamy et al. 
\cite{Veerasamy2021Bayesian} employed Naïve Bayes models to identify students 
at risk in introductory programming using assessment and self-reported cognitive 
variables. In software engineering, Shameem et al. 
\cite{SHAMEEM2023109998} and Akbar et al. \cite{AKBAR2024107352} combined 
Naïve Bayes and Logistic Regression with Genetic Algorithms to estimate project 
success probabilities under cost constraints. These studies demonstrate the use 
of posterior probability estimates for binary prediction and optimization. 
However, they focus on student performance or software-project success, whereas 
our study examines associations between stakeholder factor ratings and 
self-reported LLM familiarity and exploratory familiarity–allocation scenarios.

Across these related studies, a remaining gap is the absence of an approach
that connects a structured taxonomy of motivating and demotivating factors,
empirical multi-stakeholder evaluation, probabilistic familiarity modeling,
and exploratory allocation analysis. Although several studies evaluate
learning outcomes or LLM tool performance, they do not use stakeholder-derived
factors within a quantitative framework for comparing alternative
configurations across global and category levels under explicit modeling
assumptions. Consequently, existing research provides limited exploratory
decision support for examining how factor configurations vary across these
analytical levels.

This study addresses the identified gap through a sequential analytical design.
First, the literature-derived taxonomy \cite{khan2025integrating} provides the
nine motivating and ten demotivating factors examined in this study. Second, the
multi-stakeholder survey empirically evaluates these factors and measures
participants' self-reported LLM familiarity. Third, Naive Bayes and Logistic
Regression examine associations between the factor ratings and high self-reported
LLM familiarity. Finally, the Genetic Algorithm combines the estimated
probabilities with the defined allocation score to compare configurations through
GC, LC, and category-level
$\Delta$Fitness analyses. Thus, the descriptive, probabilistic, and allocation
analyses perform distinct but connected roles in addressing the research question.

\section{Research Methodology} \label{sec:Methodology}
To achieve the main objective of this study and address the research question formulated, the research methodology is structured into four core phases.

\begin{itemize}
    \item \textbf{Phase 1: Development of Taxonomies} -- Identification and systematic development of taxonomies for motivating and demotivating factors influencing LLM integration in software engineering education.
    
    \item \textbf{Phase 2: Questionnaire Design and Data Collection} -- Design and administration of a structured questionnaire survey to collect empirical data from relevant stakeholders.
    
    \item \textbf{Phase 3: Data Preprocessing and Probabilistic Modeling} --  Preprocessing of the survey data and development of probabilistic models to  examine associations between factor ratings and high self-reported LLM familiarity.
    
    \item \textbf{Phase 4: GA-Based Familiarity--Allocation Optimization} --
The probability estimates were combined with a simple penalty based on the 
factor levels selected by the GA. This analysis was used to compare different 
combinations of motivating and demotivating factors.
\end{itemize}
The results follow the same sequence: Phase~2 is reported in 
Sections~\ref{sec-4.1} and~\ref{sec-4.2}, Phase~3 in 
Section~\ref{sec:4.3}, and Phase~4 in Section~\ref{Sec:4.4}.

\subsection{Phase 1: Motivators and Demotivators Taxonomies} \label{Sec: MethoPhase 1: Previous Litert}
Phase~1 of this study builds directly upon our previously published literature review \cite{khan2025integrating}, which provides the conceptual foundation for conducting this empirical investigation. In that work, we performed a literature survey to identify the key factors (motivators and demotivators) influencing the integration of LLMs in software engineering education. Through thematic analysis, we systematically extracted and categorized 25 motivating factors and 30 demotivating factors affecting LLM adoption.

The identified 25 motivating factors were organized into four high-level main categories (themes): \textit{Enhancing Learning Experiences}, \textit{Assessment and Feedback in Education}, \textit{Collaboration and Peer Learning}, and \textit{Skill Development in Software Engineering Education}. Across these four categories, nine sub-themes were identified: \textit{Programming Assistance and Debugging Support}, \textit{Personalized and Adaptive Learning}, \textit{AI as a Learning Partner}, \textit{Software Engineering Process Understanding}, \textit{Conceptual Understanding and Problem Solving}, \textit{Engagement and Motivation}, \textit{Formative Feedback and Learning Support}, \textit{Automated Assessment and Grading}, and \textit{Project-Based and Inquiry-Based Learning}.

Similarly, the 30 demotivating factors were classified into four high-level main categories (themes): \textit{Learning and Educational Challenges}, \textit{Assessment and Academic Integrity}, \textit{Student Skill Development and Cognitive Load}, and \textit{Integration and Practical Implementation}. Across these categories, ten sub-themes were identified: \textit{Plagiarism and Intellectual Property Concerns}, \textit{Over-Reliance on AI in Learning}, \textit{Reduced Critical Thinking and Problem-Solving}, \textit{Ethical Concerns in AI-Assisted Learning}, \textit{Challenges in Evaluating Learning Outcomes}, \textit{Security, Privacy, and Data Integrity Issues}, \textit{Bias and Hallucination in LLM Outputs}, \textit{Limitations in Understanding and Context}, \textit{Computational and Resource Costs}, and \textit{Difficulty in Course Redesign and Curriculum Integration}. 

This thematic mapping of motivators and demotivators provides structured taxonomies that capture both the enabling conditions and the challenges associated with LLM integration in software engineering education. The detailed identification process and factor mappings are comprehensively reported in the prior review paper~\cite{khan2025integrating}. In the present study, these literature-based taxonomies serve as the theoretical backbone for survey instrument design (Phase~2) and for developing the probabilistic familiarity models and the exploratory 
familiarity--allocation framework (Phase~4), thereby extending the conceptual taxonomy toward empirical evaluation and
exploratory familiarity--allocation modeling.

\subsection{Phase 2: Designing the Survey Questionnaire}

The purpose of Phase~2 was to translate the literature-derived 
taxonomies from Phase~1 into measurable survey items and collect stakeholder data 
for the empirical and modeling phases. The questionnaire collected participants' 
background information and self-reported LLM familiarity, together with their 
ratings of the nine motivating and ten demotivating sub-themes. For each sub-theme, 
participants assessed its perceived impact and the cost or effort required either 
to implement the motivating factor or mitigate the demotivating factor. These 
assessments were recorded using a 9-point Likert scale, allowing finer 
differentiation among responses and maintaining consistency across the measurements. 
The use of a 9-point scale is also supported by previous empirical software 
engineering studies \cite{AKBAR2024107352,SHAMEEM2023109998}. Although separate 
cost--effort ratings were collected, they were not directly incorporated into the 
GA objective. Instead, the optimization used the sum of the GA-selected factor 
levels as a simplified relative allocation proxy.

The background information was summarized using frequencies 
and percentages to describe the sample and characterize the distribution of the 
participating stakeholder groups. No demographic hypotheses or inferential 
subgroup comparisons were examined. The impact ratings were used to empirically 
evaluate the literature-derived factors, while self-reported LLM familiarity was 
used as the outcome variable in the probabilistic analysis. The separate cost--effort ratings were not directly used in the optimization. 
Instead, the GA used the relative allocation penalty defined in Section~\ref{sec:3.4.2}.

\subsubsection{Piloting of the Survey Questionnaire}

Prior to full deployment, the survey questionnaire underwent a structured pretesting and validation process to ensure clarity, content validity, and alignment with the developed taxonomies as highlighted by Khan et al.\cite{khan2017systematic}. The first author initially designed the survey instrument. The second author, who possesses extensive experience in designing and conducting empirical survey studies in software engineering, subsequently reviewed the instrument to assess the appropriateness of the variables, the alignment between survey items and taxonomy factors, and the suitability of the measurement scales.

Following mutual discussion, the second author recommended restructuring the survey instrument and limiting the core items to the main categories (themes) and sub-themes defined in the taxonomies \cite{khan2025integrating}. This decision was motivated by the fact that the original list comprised 55 factors (25 motivators and 30 demotivators), which would have resulted in an excessively lengthy questionnaire and potentially reduced response quality and participation rates. In addition, careful attention was given to ensuring that each item accurately represented the intended motivating or demotivating construct and that the 9-point Likert scale provided sufficient granularity for subsequent quantitative modeling.

After this internal review and revision, the updated questionnaire was evaluated by an external expert from another university with experience in empirical software engineering research. The authors discussed the rationale for including only the main categories and sub-themes with the external evaluator and sought his feedback. The expert expressed agreement with this approach, emphasizing that survey participation is voluntary and that an overly long instrument can increase respondent burden, reduce clarity, and negatively affect completion rates. Based on the received feedback, substantial refinements were made to improve readability, reduce interpretational bias, and enhance structural coherence. 

After piloting and refinement, the survey was deployed using an online Google
Forms questionnaire. The survey invitation and questionnaire
link were distributed through the authors' academic networks. Recruitment
followed a non-probability approach combining purposive and convenience
sampling. Eligible participants included stakeholders involved in teaching,
learning, research, curriculum development, or related software engineering
education activities. No response quotas were applied. A total of 141
responses were received. After applying the exclusion criteria described in
Section \ref{Sec:Data Cleaning}, 126 complete and valid responses were retained for analysis.
The survey was conducted to empirically evaluate the
literature-derived taxonomy and collect stakeholder data for developing the
exploratory familiarity--allocation framework. It was not designed as an
institutional implementation or intervention study, and the framework was not
implemented or evaluated at the institutional level in the present study.
According to the applicable institutional guidelines in the authors'
institutional context, this survey study did not require formal ethics
committee approval. Nevertheless, the study followed standard ethical research
practices: participation was voluntary, informed consent was obtained prior to
participation, respondents could discontinue at any point, and the dataset used
for analysis did not include personally identifying information.

\subsection{Phase 3: Data Preprocessing and Probabilistic Modeling}

The purpose of Phase~3 was to prepare the survey data and 
examine the association between stakeholders' ratings of the nine motivating 
and ten demotivating sub-themes and their self-reported LLM familiarity. The 
19 factor ratings were used as predictor variables, while the 9-point 
familiarity rating was transformed into a binary outcome representing high 
familiarity versus moderate or lower familiarity. Participants' background 
information was used only to describe the sample and stakeholder groups. 
After data cleaning and encoding, Naïve Bayes and Logistic Regression models 
were trained to estimate the probability that a factor-rating configuration 
is associated with high self-reported familiarity. These probability estimates 
were subsequently used in the exploratory optimization analysis in Phase~4. 
This phase therefore consists of three steps: (i) data cleaning and validation, 
(ii) factor encoding and outcome-variable construction, and (iii) probabilistic 
model training and evaluation.

\subsubsection{Data Cleaning and Validation} \label{Sec:Data Cleaning}

From the 141 collected responses, 126 complete and valid responses were retained for analysis. Responses were excluded according to predefined criteria to ensure data quality, internal consistency, and modeling reliability, following established guidelines for empirical software engineering research presented by Kitchenham et al. \cite{kitchenham2002preliminary}. Specifically, entries were removed if:

\begin{itemize}
    \item[(i)] the respondent did not provide informed consent,
    \item[(ii)] the response pattern indicated systematic non-engagement (e.g., identical ratings across all items without variation).
\end{itemize}

After applying the two exclusion criteria listed above, 15 of the 141 responses 
were excluded, leaving $N=126$ valid responses. The dataset contained $p=9$ 
motivator variables and $q=10$ demotivator variables, giving a total of 
$d=p+q=19$ predictor variables. Each variable was encoded numerically using a 
9-point Likert scale ranging from 1 (Extremely Low) to 9 (Extremely High) 
\cite{Norman2010}.

For respondent $i$, the feature vector is defined as:

\begin{equation}
\mathbf{x}^{(i)} =
\left(x_1^{(i)},x_2^{(i)},\ldots,x_{19}^{(i)}\right),
\quad x_j^{(i)}\in\{1,2,\ldots,9\}, \quad j=1,\ldots,19.
\tag{1}
\end{equation}

Thus, the predictor matrix used for probabilistic modeling is represented as:

\begin{equation}
\mathbf{X}\in\{1,2,\ldots,9\}^{126\times19},
\tag{2}
\end{equation}

where the 126 rows represent respondents and the 19 columns represent the 
nine motivator and ten demotivator variables.

\subsubsection{Outcome Variable Construction}
To enable predictive modeling, an outcome variable reflecting the level 
of LLM familiarity was constructed. Familiarity was directly measured 
using a 9-point scale in the questionnaire. In this study, familiarity 
represents respondents' self-reported knowledge of and exposure to LLM 
technologies. It should not be interpreted as a measure of successful 
curriculum integration or pedagogical competence, as effective integration 
may also depend on curriculum design capabilities, assessment practices, 
institutional policies, infrastructure, staff training, and demonstrated 
learning outcomes.

Let the outcome variable $Y$ be defined as:
\begin{equation}
Y =
\begin{cases}
1, & \text{if familiarity rating } \geq 7, \\
0, & \text{if familiarity rating } \leq 6.
\end{cases}
\tag{3}
\end{equation}

Because no externally validated cutoff is available for this study-specific self-reported familiarity measure, we adopted a transparent scale-based operational definition. The nine ordered response categories were divided into three contiguous groups: ratings 1--3 represented low familiarity, 4--6 represented moderate familiarity, and 7--9 represented high familiarity. As the binary modeling objective was to distinguish respondents reporting high familiarity from all other respondents, the low- and moderate-familiarity groups were combined into one category. This cutoff is a study-specific operationalization of the upper range of the nine-point familiarity scale and is not presented as a universal or externally validated threshold. A binary outcome was selected because the analysis focuses on estimating membership in the high-familiarity group. Treating the measure as continuous would assume equal distances between adjacent response categories, whereas ordinal modeling would address a different analytical question involving all ordered familiarity levels.

This approach follows methodological guidance that rating-scale categories may be collapsed when the resulting categories have a meaningful interpretation and the analytical decision is explicitly justified \cite{harpe2015analyze,sullivan2013analyzing}. Accordingly, the cutoff at 7 should be understood as a transparent and reproducible study-specific operationalization, rather than an externally validated or universal threshold for LLM familiarity. This classification resulted in 90 high-familiarity cases (71.4\%) and 36 moderate-or-lower-familiarity cases (28.6\%).

The predictive modeling objective is therefore to estimate:
\begin{equation}
P(Y = 1 \mid \mathbf{x})
\tag{4}
\end{equation}
which represents the estimated probability that a given factor-rating 
configuration is associated with high self-reported LLM familiarity. 
This probability captures familiarity rather than the likelihood of 
successful or pedagogically effective LLM integration.

\subsubsection{Probabilistic Model Training and Evaluation}

Following data preprocessing and outcome construction, probabilistic classification models were trained to estimate the likelihood that a given configuration of motivating and demotivating factors corresponds to high familiarity with LLMs. The predictive task is therefore formulated as a binary classification problem where the input is the feature vector $\mathbf{x}^{(i)}$ representing the encoded survey responses and the target variable is the binary familiarity indicator $Y$ defined in Eq.~(3).

Two probabilistic classifiers, Naïve Bayes (NB) and Logistic Regression 
(LR), were selected because the binary outcome and the subsequent optimization 
phase require estimates of $P(Y=1\mid\mathbf{x})$ rather than only predicted 
class labels. NB is a generative classifier that applies Bayes' theorem to 
combine the class prior and class-conditional likelihood, whereas LR estimates 
the posterior probability directly. Their joint use provides complementary 
generative and discriminative perspectives \cite{NgJordan2002}. The relatively 
small sample of 126 observations with 19 predictors also favored transparent 
models with limited complexity. NB relies on the simplifying assumption that the predictors are conditionally 
independent given the outcome. Although previous research shows that NB can 
remain effective under some departures from this assumption 
\cite{Domingos1997}, the factor ratings in this study may be correlated. 
Therefore, NB was not used as the sole model; its probability estimates were 
combined with those from LR and interpreted as exploratory scoring functions 
for the optimization phase rather than reliable individual-level predictions.

Model performance was evaluated using accuracy on a stratified 80/20
train--test split. This single train--test split provides only one estimate of model performance, and the results may vary across alternative data partitions. It is therefore treated as a limited evaluation rather than definitive validation of predictive stability. As reported in Section~\ref{sec:4.3}, both models
performed below the majority-class baseline. Therefore, their probability
estimates are used only as exploratory relative scoring functions in
Phase~4 and should not be interpreted as reliable individual-level
predictions.

\subsection{Phase 4: GA-Based Familiarity--Allocation Optimization}
The purpose of Phase~4 is to compare different configurations of motivating 
and demotivating factors. For each configuration generated by the Genetic 
Algorithm (GA), the probabilistic models estimate the probability of high 
self-reported LLM familiarity. This probability is balanced against a simple 
allocation penalty based on the factor levels selected by the GA. The analysis 
includes: (i) defining the probabilistic models, (ii) combining their probability 
estimates, (iii) calculating the relative allocation penalty, and (iv) defining 
the GA fitness function. The results represent exploratory scenarios and should 
not be interpreted as validated recommendations for successful LLM integration.

\subsubsection{Probabilistic Prediction Models}

Two probabilistic classifiers (NBC, LR) were employed, both grounded in statistical learning theory as formalized by Hastie et al.~\cite{Hastie2009}. Both models estimate:

\begin{equation}
\hat{P}(Y = 1 \mid \mathbf{s})
\tag{5}
\end{equation}

\paragraph{NBC}
Based on Bayes’ theorem, the posterior probability is:

\begin{equation}
P(Y = 1 \mid \mathbf{s}) =
\frac{P(\mathbf{s} \mid Y = 1) P(Y = 1)}
{P(\mathbf{s})}
\tag{6}
\end{equation}
Equation (6) is the standard Bayes identity used to obtain 
the posterior probability of high self-reported familiarity for a candidate 
factor configuration. Bayes' theorem itself does not require conditional 
independence; the simplifying assumption specific to the Naïve Bayes classifier 
is introduced in Equation (7).

\begin{equation}
P(\mathbf{s} \mid Y = 1) = \prod_{j=1}^{d} P(s_j \mid Y = 1)
\tag{7}
\end{equation}

Substituting Eq.~(7) into Eq.~(6):

\begin{equation}
P(Y = 1 \mid \mathbf{s}) =
\frac{P(Y = 1)\prod_{j=1}^{d} P(s_j \mid Y = 1)}
{\sum_{y \in \{0,1\}} P(Y = y)\prod_{j=1}^{d} P(s_j \mid Y = y)}
\tag{8}
\end{equation}

The resulting estimate is denoted:

\begin{equation}
\hat{P}_{\mathrm{NBC}}(Y = 1 \mid \mathbf{s})
\tag{9}
\end{equation}

\paragraph{LR}
LR models the log-odds of the outcome as:

\begin{equation}
z = \beta_0 + \sum_{j=1}^{d} \beta_j s_j
\tag{10}
\end{equation}

The probability estimate is obtained via the logistic function:

\begin{equation}
P(Y = 1 \mid \mathbf{s}) =
\frac{1}{1 + e^{-z}} =
\frac{1}{1 + e^{-(\beta_0 + \sum_{j=1}^{d} \beta_j s_j)}}
\tag{11}
\end{equation}

The LR-based probability estimate is denoted:

\begin{equation}
\hat{P}_{\mathrm{LR}}(Y = 1 \mid \mathbf{s})
\tag{12}
\end{equation}

\paragraph{Aggregated Probability Estimation}

To reduce dependence on a single model and account for modeling uncertainty, an aggregated probability estimate is defined as:

\begin{equation}
\hat{P}_{\mathrm{agg}}(\mathbf{s}) =
\frac{
\hat{P}_{\mathrm{NBC}}(\mathbf{s}) +
\hat{P}_{\mathrm{LR}}(\mathbf{s})
}{2}
\tag{13}
\end{equation}

where $\mathbf{s}$ represents a candidate factor configuration. The averaging in Equation (13) is used only to reduce dependence on either model alone. It does not establish probability calibration or stability, particularly for synthetic candidate configurations that are weakly represented in the observed data.

\subsubsection{Allocation Score and Fitness Formulation} \label{sec:3.4.2} The allocation penalty is calculated from the factor levels in each candidate configuration. A candidate configuration is defined as:

\begin{equation}
\mathbf{s} = (s_1, s_2, \ldots, s_d),
\quad s_j \in \{1,2,\ldots,9\}
\tag{14}
\end{equation}

The allocation penalty is defined as:

\begin{equation}
C(\mathbf{s}) = \sum_{j=1}^{d} s_j
\tag{15}
\end{equation}

Here, \(C(\mathbf{s})\) is the sum of the factor levels selected by the GA. The separate cost--effort ratings collected in the survey are not used in this calculation. Therefore, \(C(\mathbf{s})\) represents a relative allocation score, not an actual institutional cost. The equal, linear formulation is used as a transparent baseline penalty for comparing configurations rather than as an empirical representation of institutional resources. The cost--effort ratings are subjective ordinal perceptions obtained from heterogeneous respondents and were not validated as comparable, institution-specific estimates of staff time, training, licensing, infrastructure, or policy effort. Consequently, using them as cardinal weights would introduce unsupported precision and a resource interpretation not warranted by the data. A higher selected level produces a higher allocation score. This score is later subtracted in the fitness function to prevent the GA from selecting the highest level for every factor.

To compare the allocation score with probability values in the range $[0,1]$, it is normalized as follows:

\begin{equation}
C_{\mathrm{norm}}(\mathbf{s}) =
\frac{C(\mathbf{s}) - C_{\min}}
{C_{\max} - C_{\min}}
\tag{16}
\end{equation}

where

\begin{equation}
C_{\min} = d \times 1,
\quad
C_{\max} = d \times 9
\tag{17}
\end{equation}

The effectiveness (fitness) function is defined as:

\begin{equation}
F(\mathbf{s}) =
\hat{P}_{\mathrm{agg}}(\mathbf{s})
-
C_{\mathrm{norm}}(\mathbf{s})
\tag{18}
\end{equation}

The optimization objective is:

\begin{equation}
\max_{\mathbf{s} \in \{1,\ldots,9\}^{d}} F(\mathbf{s})
\tag{19}
\end{equation}

This formulation combines the estimated probability of high self-reported familiarity and the normalized allocation score into a single fitness value.

\subsubsection{Genetic Algorithm (GA)}

The total search space contains $9^d$ possible configurations, which is computationally infeasible for exhaustive enumeration. Therefore, a Genetic Algorithm (GA) \cite{Goldberg1989} is employed. Each chromosome encodes a full configuration $\mathbf{s}$. The fitness of each chromosome is computed using Eq.~(18). Starting from a randomly initialized population, the GA iteratively applies selection, crossover, and mutation operators over multiple generations until a predefined stopping criterion is reached. This evolutionary search identifies a configuration that balances the estimated probability of high self-reported familiarity with the normalized allocation score.

\subsection{Replication Package and Data Availability}
\begin{sloppypar}
We prepared a replication package to support the reproducibility of the study~\cite{khan2026replication}.
It contains the raw survey export with all 141 responses (\texttt{Responses.csv}), the cleaned dataset
(\texttt{Responses\_\allowbreak Cleaned\_\allowbreak Replication.csv}), the
126-response analysis dataset
(\texttt{Responses\_\allowbreak 126\_\allowbreak Complete\_\allowbreak for\_\allowbreak Analysis.csv}),
and the Python script
(\texttt{motivators\_\allowbreak demotivators.py}) used for data preparation,
descriptive analysis, construction of the binary \texttt{Success} outcome,
probabilistic modeling, and GA-based allocation analysis. The package also
provides the global allocation results
(\texttt{best\_\allowbreak solution\_\allowbreak table\_\allowbreak global\_\allowbreak GA.csv}),
the category-level results
(\texttt{theme\_\allowbreak results\_\allowbreak summary.csv}), and the local
allocation results for each of the eight categories
(\texttt{theme\_\allowbreak best\_\allowbreak allocations\_\allowbreak *.csv}).
The analyses were conducted in Python using NumPy, Pandas, scikit-learn, and DEAP.
\end{sloppypar}
\section{Results} \label{Sec4:Results}

This section presents the empirical findings of the study. We first describe
the respondents' demographic characteristics to contextualize the survey
sample. We then report the descriptive factor ratings, probabilistic modeling
results, and GA-based allocation analyses, including GC, LC, and category-level $\Delta$Fitness.

\subsection{Respondent Characteristics} \label{sec-4.1}
This subsection reports the frequencies and percentages of the respondents' background characteristics to contextualize the survey sample; no demographic hypotheses or inferential group comparisons were examined. Overall, the sample was predominantly university-based (81.7\%) and included various academic roles, such as researchers, postdoctoral researchers, lecturers, and professors. Some respondents reported multiple responsibilities, including both teaching and research (Table~\ref{tab:demographics}). Respondents were distributed across multiple countries, with the largest groups from Finland (31.7\%), Saudi Arabia (15.1\%), Pakistan (13.5\%), and China (10.3\%), alongside respondents from Europe, Asia, and North America ($\leq 5$ from each additional country).

Most respondents rated their LLM familiarity between 7 and 9 on the 9-point scale. This concentration resulted in comparatively fewer respondents with moderate or lower familiarity and should therefore be considered when interpreting the probabilistic modeling and allocation results. Furthermore, the predominance of university-based and highly familiar respondents may limit the applicability of the findings to less-experienced stakeholders and other educational settings.

\begin{table}[!htbp]
\centering
\caption{Demographic Characteristics of Respondents ($N = 126$)}
\label{tab:demographics}
\renewcommand{\arraystretch}{1.15}
\setlength{\tabcolsep}{6pt}

\begin{tabular}{|p{4cm}|p{5cm}|c|c|}
\hline
\textbf{Category} & \textbf{Sub-category} & \textbf{Frequency} & \textbf{Percentage (\%)} \\
\hline

\textbf{Gender} & Male & 58 & 46.0 \\
\hline
 & Female & 33 & 26.2 \\
\hline
 & Prefer not to say & 30 & 23.8 \\
\hline
 & Non-binary & 5 & 4.0 \\
\hline

\textbf{Country (Top Represented)} & Finland & 40 & 31.7 \\
\hline
 & Saudi Arabia & 19 & 15.1 \\
\hline
 & Pakistan & 17 & 13.5 \\
\hline
 & China & 13 & 10.3 \\
\hline
 & UK & 6 & 4.8 \\
\hline
 & Other countries ($\leq$5 each) & 31 & 24.6 \\
\hline

\textbf{Job Title} & Researcher & 30 & 23.8 \\
\hline
 & Postdoctoral Researcher & 26 & 20.6 \\
\hline
 & Lecturer & 25 & 19.8 \\
\hline
 & Professor & 22 & 17.5 \\
\hline
 & Teaching Assistant & 13 & 10.3 \\
\hline
 & Curriculum Developer & 11 & 8.7 \\
\hline
 & Educational Developer & 9 & 7.1 \\
\hline
 & Admin & 7 & 5.6 \\
\hline
 & Master Student & 4 & 3.2 \\
\hline

\multicolumn{4}{|p{\dimexpr\textwidth-2\tabcolsep-2\arrayrulewidth}|}{\textbf{Note:} Percentages are computed using the 126 valid responses as the denominator. The \textit{Job Title} variable permitted multiple selections; consequently, category percentages may sum to more than 100\%.} \\
\hline

\textbf{Institution Type} & University & 103 & 81.7 \\
\hline
 & University of Applied Sciences & 12 & 9.5 \\
\hline
 & Polytechnic & 10 & 7.9 \\
\hline
 & College & 1 & 0.8 \\
\hline

\textbf{Familiarity with LLMs (7--9 scale dominant)} & 7 & 44 & 34.9 \\
\hline
 & 8 & 40 & 31.7 \\
\hline
 & 9 & 6 & 4.8 \\
\hline
 & $\leq 6$ & 36 & 28.6 \\
\hline

\end{tabular}
\end{table}

\subsection{Descriptive Analysis} \label{sec-4.2}

Following the demographic overview, we next examine the central tendencies and dispersion of the identified motivator and demotivator factors. This descriptive analysis serves two important purposes. First, it provides an empirical grounding for understanding how stakeholders in software engineering education perceive the relative importance of motivating factors and severity of each demotivating factor. By examining the mean (M) and standard deviation (SD) of each factor, we can identify which motivating factors are perceived as comparatively strong drivers of AI adoption and which demotivating factors are considered substantial barriers, as well as assess the degree of consensus or variability among respondents \cite{wohlin2012experimentation}. 

All motivator and demotivator factors were measured on a 9-point Likert scale (1 = extremely low, 9 = extremely high). Likert-type scales with multiple response categories can be analyzed using parametric descriptive statistics such as mean and standard deviation when treated as approximately interval data \cite{Norman2010}. Tables~\ref{tab:motivators} and~\ref{tab:demotivators} report the mean, standard deviation, and number of observations (\textit{N}) for each motivating and demotivating factor, respectively, calculated from the 126 valid responses retained for analysis.

\begin{table}[!htbp]
\centering
\caption{Descriptive statistics for motivators ($N = 126$)}
\label{tab:motivators}
\renewcommand{\arraystretch}{1.15}
\setlength{\tabcolsep}{6pt}

\begin{tabular}{|p{8.2cm}|c|c|}
\hline
\textbf{Motivating Factors} & \textbf{Mean} & \textbf{SD} \\
\hline
Programming Assistance and Debugging Support & 5.261 & 1.272 \\
\hline
Personalized and Adaptive Learning & 5.047 & 1.349 \\
\hline
AI as a Learning Partner & 5.023 & 1.411 \\
\hline
Software Engineering Process Understanding & 5.007 & 1.353 \\
\hline
Conceptual Understanding and Problem Solving & 4.888 & 1.415 \\
\hline
Engagement and Motivation & 4.881 & 1.542 \\
\hline
Formative Feedback and Learning Support & 4.746 & 1.528 \\
\hline
Automated Assessment and Grading & 4.738 & 1.529 \\
\hline
Project-Based and Inquiry-Based Learning & 4.563 & 1.597 \\
\hline
\end{tabular}
\end{table}

\begin{itemize}
    \item \textit{Motivators:} We interpret the descriptive statistics by examining both the relative ranking of the mean values and their associated standard deviations to assess not only the perceived strength of each motivator but also the level of agreement among respondents. Based on the results presented in Table~\ref{tab:motivators}, the highest-rated motivator is \textit{Programming Assistance and Debugging Support} (\textit{M = 5.261}). This indicates that respondents most strongly associate LLM integration with practical support for programming tasks and error resolution. Prior empirical studies on AI-assisted coding tools report similar findings, showing that such systems can support debugging, code completion, and productivity improvement in programming tasks \cite{Vaithilingam2022,Barke2023}. In software engineering education, this suggests that LLMs are perceived as particularly valuable when they reduce hands-on friction during coding activities, such as debugging, interpreting compiler or runtime errors, and generating test scaffolds.

    The next highest-rated motivator is \textit{Personalized and Adaptive Learning} (\textit{M = 5.047}), followed closely by \textit{AI as a Learning Partner} (\textit{M = 5.023}). These results indicate that stakeholders perceive LLMs not only as productivity tools but also as interactive systems capable of providing tailored explanations, adaptive feedback, and ongoing learning support. This interpretation is consistent with recent discussions on generative AI in education, which emphasize conversational tutoring and personalized guidance as key strengths of LLM-based systems \cite{Kasneci2023}. Notably, both of these motivators rank above \textit{Automated Assessment and Grading} (\textit{M = 4.738}), suggesting that respondents attribute greater immediate educational value to interactive and adaptive support than to assessment automation.

    Overall, the \textit{SD} values range between approximately \textit{1.27} and \textit{1.60} on the 9-point scale, indicating moderate dispersion. Given the width of the scale, these values suggest a reasonable level of agreement rather than strong polarization. While perceptions are not completely uniform, the variability does not alter the overall ranking of motivators. Instead, the findings indicate general convergence in viewing LLMs as beneficial, particularly for programming assistance and adaptive learning support.
\end{itemize}

\begin{itemize}
    \item \textit{Demotivators}: Similar to motivators, we computed the mean and standard deviation values to identify which demotivating factors represent the most salient challenges to LLM adoption in software engineering education. Based on the results presented in Table~\ref{tab:demotivators}, concerns related to academic integrity and learning dependency emerge as the most prominent perceived barriers. The highest-rated demotivator is \textit{Plagiarism and Intellectual Property Concerns} (\textit{M = 5.373}). This indicates that respondents view authorship ambiguity and misuse risks as the primary institutional challenge associated with AI-assisted learning. Such concerns are widely reported in recent study conducted by Cotton et al.\cite{cotton2024} on generative AI in higher education, particularly regarding academic integrity and authorship transparency.

    Closely following are \textit{Over-Reliance on AI in Learning} (\textit{M = 5.265}) and \textit{Reduced Critical Thinking and Problem-Solving} (\textit{M = 5.163}). Although slightly lower than plagiarism concerns, these results reinforce the perception that the central risk is pedagogical rather than technical. Respondents appear concerned that students may delegate cognitive effort to LLMs, potentially weakening independent reasoning and deep conceptual engagement. Similar risks regarding over-dependence and reduced critical reflection have been highlighted by Kasneci et al.\cite{Kasneci2023} in recent analyses of generative AI in education .

    Integrity and reliability-related factors also receive elevated mean scores. \textit{Security, Privacy, and Data Integrity Issues} (\textit{M = 5.063}) and \textit{Bias and Hallucination in LLM Outputs} (\textit{M = 5.017}) are both rated above the neutral midpoint, indicating sustained caution regarding governance, transparency, and output reliability. Broader evaluations of large language models similarly emphasize bias, hallucination, and trustworthiness as key challenges for responsible deployment \cite{Dwivedi2023}. These results suggest that, while LLMs may offer pedagogical benefits, institutional acceptance depends heavily on safeguards that ensure responsible use and output reliability.

    In contrast, structural and operational barriers such as \textit{Computational and Resource Costs} (\textit{M = 4.825}) and \textit{Difficulty in Course Redesign and Curriculum Integration} (\textit{M = 4.603}) receive comparatively lower mean ratings. This ordering suggests that respondents do not primarily perceive LLM adoption as constrained by infrastructure or financial feasibility. Rather, resistance appears to be anchored in concerns about learning quality, integrity, and cognitive development. From a managerial perspective, this implies that infrastructure investments alone are unlikely to ensure successful adoption without parallel efforts in assessment redesign, policy development, and pedagogical alignment.

    Regarding dispersion, standard deviations range approximately between \textit{1.20} and \textit{1.58} on the 9-point scale, indicating moderate variability without extreme polarization. This suggests a reasonable degree of agreement across respondents regarding the relative seriousness of these demotivating factors. The consistently elevated means across integrity-related concerns strengthen the robustness of this pattern and underscore that academic governance—rather than technical capability—emerges as the dominant theme shaping stakeholder perceptions.
\end{itemize}

\begin{table}[!htbp]
\centering
\caption{Descriptive statistics for demotivators ($N = 126$)}
\label{tab:demotivators}
\renewcommand{\arraystretch}{1.15}
\setlength{\tabcolsep}{6pt}

\begin{tabular}{|p{8.2cm}|c|c|}
\hline
\textbf{Demotivating Factors} & \textbf{Mean} & \textbf{SD} \\
\hline
Plagiarism and Intellectual Property Concerns & 5.373 & 1.312 \\
\hline
Over-Reliance on AI in Learning & 5.265 & 1.197 \\
\hline
Reduced Critical Thinking and Problem-Solving & 5.163 & 1.442 \\
\hline
Ethical Concerns in AI-Assisted Learning & 5.079 & 1.371 \\
\hline
Challenges in Evaluating Learning Outcomes & 5.071 & 1.415 \\
\hline
Security, Privacy, and Data Integrity Issues & 5.063 & 1.372 \\
\hline
Bias and Hallucination in LLM Outputs & 5.017 & 1.439 \\
\hline
Limitations in Understanding and Context & 4.984 & 1.379 \\
\hline
Computational and Resource Costs & 4.825 & 1.497 \\
\hline
Difficulty in Course Redesign and Curriculum Integration & 4.603 & 1.580 \\
\hline
\end{tabular}
\end{table}

\subsection{Probabilistic Modeling}
\label{sec:4.3}

Following the descriptive analysis, we examined whether
respondents' ratings of the nine motivating and ten demotivating factors
were associated with their self-reported LLM familiarity. This analysis
had two objectives: first, to assess whether the fitted models could
distinguish respondents with higher familiarity from those with moderate
or lower familiarity; and second, to construct an exploratory probabilistic
mapping for the subsequent allocation analysis
(Section~\ref{Sec:4.4}). LLM familiarity refers to respondents'
self-reported knowledge of and exposure to LLM technologies. It does not
measure successful curriculum integration, institutional readiness,
pedagogical competence, or educational effectiveness.

Using the 126 valid responses, we constructed a feature
matrix containing the 19 motivating and demotivating factors. Each
observation represents a 19-dimensional rating vector measured on the
9-point scale. Formally, the feature matrix is represented as:

\[
\mathbf{X} \in \mathbb{R}^{126 \times 19}.
\]

The self-reported familiarity rating was transformed into
a binary outcome representing high familiarity versus moderate or lower
familiarity. The resulting distribution contained 90 higher-familiarity
cases and 36 moderate- or lower-familiarity cases. This imbalance is
important because predicting the majority class for every respondent
would produce an accuracy of approximately 0.71. Familiarity was used
because it was directly measured in the questionnaire; however, it does
not demonstrate that a respondent can effectively integrate LLMs into a
course or curriculum.

Two probabilistic classifiers were trained: Naive Bayes
(NB) and Logistic Regression (LR). These models were selected because the
subsequent GA-based analysis requires continuous estimates of
$\hat{P}(Y=1\mid\mathbf{X})$, rather than only predicted class labels.
Their use also provides complementary generative and discriminative model
perspectives. The theoretical differences between NB and LR are discussed
by Ng and Jordan~\cite{NgJordan2002}, while the behavior of NB under
violations of its conditional-independence assumption is examined by
Domingos and Pazzani~\cite{Domingos1997}. Nevertheless, using two models
does not eliminate their respective assumptions or guarantee reliable
probability estimates.

An 80/20 stratified train--test split was used to preserve
the class proportions. On the held-out test set, both NB and LR achieved
an accuracy of 0.461. This value is below the approximately 0.71
majority-class baseline derived from the 90/36 class distribution.
Therefore, the fitted models did not reliably distinguish respondents
with higher familiarity from those with moderate or lower familiarity
using the 19 factor ratings. The result should not be interpreted as
evidence that the two familiarity groups necessarily have identical
perceptions; rather, it indicates that the present data, predictors, and
models provide limited discriminatory evidence.

This limited performance constrains the interpretation of
the subsequent analysis. Classification accuracy evaluates whether the
models correctly assign individual respondents to the two familiarity
classes at a fixed decision threshold. The GA does not use these binary
class assignments; it uses the continuous probability estimates as
relative scoring functions for comparing candidate factor configurations.
Consequently, the probability estimates should not be treated as reliable
individual-level predictions or as evidence that changing a factor level
will improve LLM familiarity or integration. They provide only an
exploratory probability surface under the fitted models.

During experimentation, NB occasionally produced extreme
probability estimates for synthetic configurations generated by the GA,
particularly for configurations outside densely represented regions of
the observed data. LR produced comparatively less extreme estimates across
the candidate factor space. To reduce reliance on either model alone, their
estimated probabilities were aggregated using simple averaging:

\begin{equation}
\hat{P}_{\mathrm{agg}}(\mathbf{S}) =
\frac{\hat{P}_{\mathrm{NB}}(\mathbf{S})+
\hat{P}_{\mathrm{LR}}(\mathbf{S})}{2}.
\tag{20}
\end{equation}

The resulting mapping,
\[
\mathbf{S}\rightarrow\hat{P}_{\mathrm{agg}}(\mathbf{S}),
\]
provides a continuous relative scoring function for candidate factor
configurations. This function allows the GA to compare configurations
consistently under the same fitted models and allocation assumptions.
However, averaging the estimates does not overcome the models' limited
classification performance. Therefore, the resulting allocations are
interpreted as exploratory comparative scenarios rather than externally
validated predictions, causal factor priorities, or prescriptive
institutional recommendations.

\subsection{GA-Based Allocation Analysis}
\label{Sec:4.4}
Building on the exploratory probabilistic mapping established in
Section~\ref{sec:4.3}, we conducted a three-stage GA-based allocation
analysis. First, Global Factor Allocation (GC) was performed across all
19 factors simultaneously (nine motivators and ten demotivators) to compare
factor configurations under the fitted probability models and the defined
allocation score
(Section~\ref{sec:GlobalGA_Factors}, Table~\ref{tab:global_opt}).
Second, category-level analysis was conducted across the eight higher-level
categories to calculate the change in fitness relative to each category's
baseline configuration
(Section~\ref{sec:GA_MainCategories}, Table~\ref{tab:category_opt}).
Third, Local Factor Allocation (LC) was performed independently within each
category to examine the configurations of its constituent factors
(Section~\ref{sec:GA_Within_MainCat},
Tables~\ref{tab:local_opt_mot} and~\ref{tab:local_opt_dem}).
Because the probabilistic outcome represents high self-reported LLM
familiarity, the GA-generated configurations are interpreted as exploratory
familiarity--allocation scenarios. The allocation score is a simplified,
relative proxy and does not represent actual institutional expenditure.
Therefore, the configurations do not establish causal factor priorities,
pedagogical effectiveness, institutional readiness, or validated strategies
for successful LLM integration. The GA searches the full synthetic 1--9 factor space and is not restricted to configurations close to the observed response patterns. Consequently, some candidate configurations may be weakly supported by the available data, further limiting their interpretation beyond the specified models and assumptions.
Together, these analyses provide a hierarchical comparison at three levels:
GC compares configurations across all 19 factors, category-level
$\Delta$Fitness measures improvement in the defined model objective relative
to the corresponding baseline, and LC examines configurations within each
category. All results are interpreted within the limitations of the fitted
probability models and the defined allocation score. Because the probability estimates have limited discriminatory performance, the GA is retained only as a methodological demonstration of how the specified familiarity–allocation objective behaves under the fitted models. It should not be interpreted as a validated optimizer, and the generated configurations do not provide reliable or generalizable implementation priorities.

\subsubsection{Global GA Optimization Across Factors (motivators, demotivators)} 
\label{sec:GlobalGA_Factors}

Based on Section~\ref{sec:4.3}, we used the aggregated probability 
function $\hat{P}_{\text{agg}}(\mathbf{S})$ in the GA-based 
familiarity--allocation analysis introduced in Phase~4. GA was originally proposed for solving combinatorial and resource-allocation problems under competing objectives \cite{Goldberg1989}. The objective of this stage is to identify configurations that balance the estimated probability of high self-reported LLM familiarity with the allocation score.

The GA searches across integer-encoded configurations \(\mathbf{S}\), where 
each of the 19 factors is assigned a level from 1 to 9. Each candidate 
configuration is therefore a vector of 19 integers. A higher value represents 
a higher modeled factor level, while a lower value represents a lower modeled 
factor level. These values are used to calculate the allocation score and do 
not represent actual institutional effort or cost.

\begin{table}[!htbp]
\centering
\caption{Global GA Optimization Results Across All Factors}
\label{tab:global_opt}
\renewcommand{\arraystretch}{1.1}
\setlength{\tabcolsep}{4pt}

\resizebox{\columnwidth}{!}{%
\begin{tabular}{|p{0.72\columnwidth}|c|c|}
\hline
\textbf{Factors} & \textbf{Selected Level (1--9)} & \textbf{Allocation Score} \\
\hline
Plagiarism and Intellectual Property Concerns & 8 & 8 \\
\hline
Ethical Concerns in AI-Assisted Learning & 6 & 6 \\
\hline
Automated Assessment and Grading & 4 & 4 \\
\hline
Computational and Resource Costs & 4 & 4 \\
\hline
Bias and Hallucination in LLM Outputs & 2 & 2 \\
\hline
Engagement and Motivation & 2 & 2 \\
\hline
Difficulty in Course Redesign and Curriculum Integration & 2 & 2 \\
\hline
AI as a Learning Partner & 1 & 1 \\
\hline
Formative Feedback and Learning Support & 1 & 1 \\
\hline
Over-Reliance on AI in Learning & 1 & 1 \\
\hline
Limitations in Understanding and Context & 1 & 1 \\
\hline
Reduced Critical Thinking and Problem-Solving & 1 & 1 \\
\hline
Personalized and Adaptive Learning & 1 & 1 \\
\hline
Challenges in Evaluating Learning Outcomes & 1 & 1 \\
\hline
Programming Assistance and Debugging Support & 1 & 1 \\
\hline
Project-Based and Inquiry-Based Learning & 1 & 1 \\
\hline
Conceptual Understanding and Problem Solving & 1 & 1 \\
\hline
Security, Privacy, and Data Integrity Issues & 1 & 1 \\
\hline
Software Engineering Process Understanding & 1 & 1 \\
\hline
\end{tabular}%
}
\end{table}

In this study, the allocation score for each factor is equal to its selected 
level. The total allocation score is the sum of the selected levels across all 
19 factors. Therefore, higher factor levels produce a higher allocation score, 
while lower levels produce a lower score. The allocation score is normalized 
to the range $[0,1]$ as \(C_{\mathrm{norm}}(\mathbf{S})\), using the formulation 
defined in Section~\ref{sec:3.4.2}. The fitness function is defined as:

\begin{equation}
F(\mathbf{S})=\hat{P}_{\text{agg}}(\mathbf{S})-C_{\text{norm}}(\mathbf{S})
\tag{21}
\end{equation}

where \(\hat{P}_{\text{agg}}(\mathbf{S})\) is the aggregated probability 
from NB and LR (Section~\ref{sec:4.3}), and 
\(C_{\mathrm{norm}}(\mathbf{S})\) is the normalized allocation score. 
The fitness value increases when the estimated probability is higher and 
the allocation score is lower. Similar trade-off-based formulations are 
commonly used in evolutionary optimization \cite{Deb2001}.

To evaluate the optimization gain, we defined a global baseline configuration
by rounding the sample mean rating of each of the 19 factors to the nearest
integer level. The resulting baseline configuration had a total allocation
score of 135 and an aggregated probability of 0.769. Using the normalization
scheme defined in Section~\ref{sec:3.4.2}, with \(d=19\),
\(C_{\min}=19\), and \(C_{\max}=171\), its normalized allocation score was:

\[
C_{\mathrm{norm}}(\mathbf{S}_{\mathrm{base}})
=
\frac{135-19}{171-19}
=
0.763.
\]

Accordingly, the global baseline fitness was:

\[
F(\mathbf{S}_{\mathrm{base}})
=
0.769-0.763
=
0.006.
\]

Thus, the baseline configuration had a fitness value close to zero because
its comparatively high normalized allocation score largely offset its
aggregated probability.

After 40 generations, the GA produced a configuration with a higher fitness
than the baseline. Under the defined fitness function, this improvement was
mainly attributable to the substantially lower normalized allocation score of
the GA-generated configuration. Therefore, the identified configuration
achieved a better familiarity--allocation balance than the baseline under the
fitted models and the defined allocation score.
The selected levels in Table~\ref{tab:global_opt} produced a total
allocation score of 40. Using the normalization scheme defined in
Section~\ref{sec:3.4.2}, the normalized allocation score of the
GA-generated configuration was:

\[
C_{\mathrm{norm}}(\mathbf{S}_{\mathrm{GA}})
=
\frac{40-19}{171-19}
=
0.138.
\]
Inspection of the global allocation results (Table~\ref{tab:global_opt}) shows 
that \textit{Plagiarism and Intellectual Property Concerns} (level 8) and 
\textit{Ethical Concerns in AI-Assisted Learning} (level 6) received the highest 
selected levels. Under the fitted probability models and the defined allocation 
score, the GA favored these integrity- and governance-related factors in the 
identified configuration.

\textit{Automated Assessment and Grading} and \textit{Computational and 
Resource Costs} received intermediate levels (4). \textit{Bias and 
Hallucination in LLM Outputs}, \textit{Engagement and Motivation}, and 
\textit{Difficulty in Course Redesign and Curriculum Integration} received 
lower levels (2). Most remaining factors received level 1. These lower levels 
do not mean that the factors have little educational value. They only reflect 
the balance produced by the fitted probability models and the defined 
allocation score.

Overall, the global analysis shows that the baseline configuration has a less favorable familiarity--allocation balance under the fitted models and the defined 
allocation score. The GA-generated configuration assigns relatively higher levels 
to governance- and integrity-related factors than to several pedagogical factors. 
However, this is a model-based result. It does not demonstrate that these factors 
cause successful LLM integration or should always receive priority in practice.

\begin{tcolorbox}[
colback=gray!12,
colframe=gray!45,
arc=3mm,
boxrule=0.8pt,
width=\columnwidth,
enhanced,
breakable,
title={\faKey\hspace{0.5em}Key Insights from Factor-level Global GA Optimization},
fonttitle=\bfseries,
coltitle=black,
boxed title style={
    colback=gray!20,
    colframe=gray!45,
    arc=2mm
},
attach boxed title to top left={xshift=4mm,yshift=-2mm},
]
Under the fitted probability models and the defined allocation
score, the GA-generated configuration assigns relatively high levels to
governance- and integrity-related factors, particularly plagiarism management
and ethical AI practices. These allocations represent exploratory
familiarity--allocation scenarios rather than evidence that these factors cause
successful integration. Institutional and classroom-based studies are required
before using these results as implementation recommendations.

\end{tcolorbox}

\subsubsection{Category-Level GA Optimization} 
\label{sec:GA_MainCategories}

While Section~\ref{sec:GlobalGA_Factors} analyzes all 19 motivating and
demotivating factors together, this subsection examines the eight main
categories separately. The purpose is to compare the balance between the
estimated probability of high self-reported LLM familiarity and the normalized
allocation score across these categories.

Table~\ref{tab:category_opt} presents four measures for each category.
GA\_P\_Agg is the estimated probability of high self-reported LLM familiarity
for the GA-generated configuration. GA\_NormAllocation is the normalized
allocation score. GA\_Fitness represents the balance between these two values
and is calculated by subtracting the normalized allocation score from the
estimated probability. Finally, $\Delta$\_Fitness represents the improvement
in fitness compared with the baseline configuration. A larger
$\Delta$\_Fitness indicates a greater improvement under the defined fitness
function within that category.

\begin{table}[!htbp] \centering \caption{Main Categories-Level GA Results} \label{tab:category_opt} \renewcommand{\arraystretch}{1.15} \setlength{\tabcolsep}{3pt} \resizebox{\columnwidth}{!}{\begin{tabular}{|>{\raggedright\arraybackslash}p{0.42\columnwidth}|c|c|c|c|c|} \hline \textbf{Theme} & \textbf{\shortstack{Num\_\\Factors}} & \textbf{\shortstack{GA\_P\_Ag\\g}} & \textbf{\shortstack{GA\_NormCos\\t}} & \textbf{\shortstack{GA\_Fitnes\\s}} & \textbf{\shortstack{$\Delta$\\Fitness}} \\ \hline MC4\_Collaboration and Peer Learning & 2 & 0.976 & 0.463 & 0.513 & 0.244 \\ \hline MC2\_Enhancing Learning Experiences & 3 & 0.936 & 0.463 & 0.473 & 0.203 \\ \hline MC3\_Assessment and Feedback in Education & 2 & 0.970 & 0.513 & 0.456 & 0.187 \\ \hline MC1\_Skill Development in Software Engineering Education & 2 & 0.964 & 0.527 & 0.436 & 0.167 \\ \hline DC4\_Integration and Practical Implementation & 3 & 0.933 & 0.504 & 0.428 & 0.159 \\ \hline DC2\_Learning and Educational Challenges & 3 & 0.928 & 0.509 & 0.419 & 0.150 \\ \hline DC3\_Student Skill Development and Cognitive Load & 2 & 0.876 & 0.486 & 0.390 & 0.120 \\ \hline DC1\_Assessment and Academic Integrity & 2 & 0.877 & 0.532 & 0.344 & 0.075 \\ \hline \multicolumn{6}{|>{\raggedright\arraybackslash}p{0.98\columnwidth}|}{MC1--MC4 (Motivators Main Categories/themes), DC1--DC4 (Demotivators Main Categories/themes)} \\ \hline
\multicolumn{6}{l}{\footnotesize All values are rounded to three decimal 
places; fitness values were calculated using unrounded values.}
\end{tabular} } \end{table}

The largest improvement is observed for \textit{MC4\_Collaboration and Peer Learning} 
(\(\Delta\)\_Fitness = 0.244), followed by 
\textit{MC2\_Enhancing Learning Experiences} 
(\(\Delta\)\_Fitness = 0.203). Both categories have a normalized allocation 
score of 0.463. Their relatively high estimated probabilities and lower 
normalized allocation scores produce the two largest improvements under the 
defined fitness function.

The improvements for \textit{MC3\_Assessment and Feedback in Education}
and \textit{MC1\_Skill Development in Software Engineering Education}
are 0.187 and 0.167, respectively. Although both categories have high
estimated probabilities, their higher normalized allocation scores
(0.513 and 0.527) reduce their fitness improvements.

Among the demotivating categories,
\textit{DC4\_Integration and Practical Implementation} has the largest
improvement (\(\Delta\)\_Fitness = 0.159), with a normalized allocation
score of 0.504. It is followed by
\textit{DC2\_Learning and Educational Challenges}
(\(\Delta\)\_Fitness = 0.150, GA\_NormAllocation = 0.509) and
\textit{DC3\_Student Skill Development and Cognitive Load}
(\(\Delta\)\_Fitness = 0.120). 
\textit{DC1\_Assessment and Academic Integrity} has the smallest
improvement (\(\Delta\)\_Fitness = 0.075) and the highest normalized
allocation score (0.532).

Overall, the category-level analysis shows that the fitness function responds
differently across the eight categories. The two pedagogically oriented
categories, \textit{Collaboration and Peer Learning} and
\textit{Enhancing Learning Experiences}, show the largest improvements under
the fitted probability models and the defined allocation score. In comparison,
the demotivating categories generally show smaller improvements. These results
represent model-based familiarity--allocation differences and do not directly
measure classroom effectiveness, actual institutional cost, or causal effects.

\subsubsection{GA Optimization Across Motivators and Demotivators within Main Categories} 
\label{sec:GA_Within_MainCat}

While Section~\ref{sec:GA_MainCategories} examined the
familiarity--allocation balance at the category level, this subsection
analyzes the factors within each category. The purpose is to examine how
the GA-selected levels differ among the specific motivators and
demotivators. Tables~\ref{tab:local_opt_mot}
and~\ref{tab:local_opt_dem} report the selected level
(\textit{Best\_Level}, 1--9) for each factor. Higher selected levels
produce higher allocation scores, but they do not represent actual
institutional costs, implementation effort, or validated recommendations.
\begin{table}[!htbp]
\centering
\caption{Local GA Results within Main Categories (Motivators)}
\label{tab:local_opt_mot}
\setlength{\tabcolsep}{2.5pt}       
\small

\begin{tabularx}{\columnwidth}{|>{\raggedright\arraybackslash}p{0.30\columnwidth}|
                                >{\raggedright\arraybackslash}X|
                                c|c|}
\hline
\textbf{Theme} & \textbf{Factor} & \textbf{\shortstack{Best Level (1--9)}} & \textbf{Allocation Score} \\
\hline

\multirow{2}{=}{MC1\_Skill Development in Software Engineering Education}
& Software Engineering Process Understanding & 9 & 9 \\
\cline{2-4}
& Programming Assistance and Debugging Support & 4 & 4 \\
\hline

\multirow{3}{=}{MC2\_Enhancing Learning Experiences}
& Conceptual Understanding and Problem Solving & 9 & 9 \\
\cline{2-4}
& Personalized and Adaptive Learning & 3 & 3 \\
\cline{2-4}
& Engagement and Motivation & 1 & 1 \\
\hline

\multirow{2}{=}{MC3\_Assessment and Feedback in Education}
& Formative Feedback and Learning Support & 9 & 9 \\
\cline{2-4}
& Automated Assessment and Grading & 7 & 7 \\
\hline

\multirow{2}{=}{MC4\_Collaboration and Peer Learning}
& AI as a Learning Partner & 5 & 5 \\
\cline{2-4}
& Project-Based and Inquiry-Based Learning & 1 & 1 \\
\hline

\multicolumn{4}{|l|}{\small MC1--MC4 (Motivators Main Categories/themes)} \\
\hline

\end{tabularx}
\end{table}

\begin{itemize}

\item \textit{Motivators Main Categories}

For the motivator categories, the GA selected different factor levels across the
four pedagogical categories (Table~\ref{tab:local_opt_mot}). These levels reflect
the familiarity--allocation balance produced by the fitted probability models and
the defined allocation score.

Within \textit{MC1\_Skill Development in Software Engineering Education},
the GA selected level 9 for \textit{Software Engineering Process
Understanding} and level 4 for \textit{Programming Assistance and Debugging
Support}. Under the fitted probability models and the defined allocation
score, process understanding therefore received the higher modeled level
within this category. This result does not demonstrate that it is
educationally more effective than programming assistance.

Within \textit{MC2\_Enhancing Learning Experiences}, the GA selected level 9
for \textit{Conceptual Understanding and Problem Solving}, level 3 for
\textit{Personalized and Adaptive Learning}, and level 1 for
\textit{Engagement and Motivation}. Under the fitted probability models and
the defined allocation score, conceptual understanding therefore received
the highest modeled level within this category. These results reflect an
exploratory familiarity--allocation scenario and do not establish the
educational effectiveness of these factors.

For \textit{MC3\_Assessment and Feedback in Education}, the GA selected
level 9 for \textit{Formative Feedback and Learning Support} and level 7
for \textit{Automated Assessment and Grading}. Under the fitted probability
models and the defined allocation score, formative feedback therefore
received the higher modeled level within this category. This result does
not establish that formative feedback is educationally more effective
than automated assessment.

Within \textit{MC4\_Collaboration and Peer Learning}, the GA selected
level 5 for \textit{AI as a Learning Partner} and level 1 for
\textit{Project-Based and Inquiry-Based Learning}. Under the fitted
probability models and the defined allocation score, AI as a learning
partner therefore received the higher modeled level within this
category. This result does not establish that it is educationally more
effective than project-based and inquiry-based learning.

\item \textit{Demotivators Main Categories}
For the demotivator categories, the GA selected different factor levels across
the four categories (Table~\ref{tab:local_opt_dem}). These levels reflect the
familiarity--allocation balance produced by the fitted probability models and
the defined allocation score. Higher selected levels do not necessarily mean
that the corresponding challenges are more serious or important in practice.

Within \textit{DC1\_Assessment and Academic Integrity}, the GA selected
level 8 for \textit{Ethical Concerns in AI-Assisted Learning} and level 6
for \textit{Plagiarism and Intellectual Property Concerns}. Under the
fitted probability models and the defined allocation score, ethical
concerns therefore received the higher modeled level within this category.
This result does not establish that ethical concerns are more important
than plagiarism and intellectual property concerns in practice.

Within \textit{DC2\_Learning and Educational Challenges}, the GA selected
level 8 for \textit{Bias and Hallucination in LLM Outputs}, level 6 for
\textit{Limitations in Understanding and Context}, and level 2 for
\textit{Over-Reliance on AI in Learning}. Under the fitted probability
models and the defined allocation score, bias and hallucination therefore
received the highest modeled level within this category. These selected
levels do not establish the relative seriousness or educational impact
of these challenges in practice.

For \textit{DC3\_Student Skill Development and Cognitive Load}, the GA
selected level 7 for \textit{Reduced Critical Thinking and Problem-Solving}
and level 2 for \textit{Challenges in Evaluating Learning Outcomes}. Under
the fitted probability models and the defined allocation score, reduced
critical thinking and problem-solving therefore received the higher modeled
level within this category. These selected levels do not establish the
relative seriousness or educational impact of these challenges in practice.

Finally, within \textit{DC4\_Integration and Practical Implementation}, the GA
selected level 9 for \textit{Security, Privacy, and Data Integrity Issues},
level 8 for \textit{Computational and Resource Costs}, and level 4 for
\textit{Difficulty in Course Redesign and Curriculum Integration}. Under the
fitted probability models and the defined allocation score, security, privacy,
and data integrity therefore received the highest modeled level within this
category. These selected levels do not establish the relative importance or
actual implementation cost of these challenges in practice.

\end{itemize}

Taken together, the factor-level results show that the GA selected different
levels across the motivator factors. \textit{Conceptual Understanding and
Problem Solving}, \textit{Software Engineering Process Understanding}, and
\textit{Formative Feedback and Learning Support} received the highest selected
levels. In contrast, \textit{Engagement and Motivation},
\textit{Project-Based and Inquiry-Based Learning}, and several other factors
received lower levels. These differences reflect the familiarity--allocation
balance produced by the fitted probability models and the defined allocation
score; they do not establish the relative educational effectiveness of the
factors.

On the demotivator side, \textit{Security, Privacy, and Data Integrity Issues}
received the highest selected level (9). \textit{Ethical Concerns in
AI-Assisted Learning}, \textit{Bias and Hallucination in LLM Outputs}, and
\textit{Computational and Resource Costs} received level 8, while
\textit{Reduced Critical Thinking and Problem-Solving} received level 7.
Other demotivator factors received lower levels. These differences reflect
the familiarity--allocation balance produced by the fitted probability
models and the defined allocation score; they do not establish the relative
seriousness or practical importance of these challenges.

Overall, the combined results in
Section~\ref{sec:GA_Within_MainCat} show that the GA selected different
factor levels under the fitted probability models and the defined allocation
score. The resulting configurations represent exploratory
familiarity--allocation scenarios rather than evidence of effective LLM
integration or recommended institutional priorities. Their practical and
educational effectiveness requires validation through future classroom and
institutional studies.

\begin{table}[!htbp]
\centering
\caption{Local GA Results within Main Categories (Demotivators)}
\label{tab:local_opt_dem}
\renewcommand{\arraystretch}{1.0}   
\setlength{\tabcolsep}{2.5pt}       
\small

\begin{tabularx}{\columnwidth}{|>{\raggedright\arraybackslash}p{0.30\columnwidth}|
                                >{\raggedright\arraybackslash}X|
                                c|c|}
\hline
\textbf{Theme} & \textbf{Factor} & \textbf{\shortstack{Best Level (1--9)}} & \textbf{Allocation Score} \\
\hline

\multirow{2}{=}{DC1\_Assessment and Academic Integrity}
& Ethical Concerns in AI-Assisted Learning & 8 & 8 \\
\cline{2-4}
& Plagiarism and Intellectual Property Concerns & 6 & 6 \\
\hline

\multirow{3}{=}{DC2\_Learning and Educational Challenges}
& Bias and Hallucination in LLM Outputs & 8 & 8 \\
\cline{2-4}
& Limitations in Understanding and Context & 6 & 6 \\
\cline{2-4}
& Over-Reliance on AI in Learning & 2 & 2 \\
\hline

\multirow{2}{=}{DC3\_Student Skill Development and Cognitive Load}
& Reduced Critical Thinking and Problem-Solving & 7 & 7 \\
\cline{2-4}
& Challenges in Evaluating Learning Outcomes & 2 & 2 \\
\hline

\multirow{3}{=}{DC4\_Integration and Practical Implementation}
& Security, Privacy, and Data Integrity Issues & 9 & 9 \\
\cline{2-4}
& Computational and Resource Costs & 8 & 8 \\
\cline{2-4}
& Difficulty in Course Redesign and Curriculum Integration & 4 & 4 \\
\hline

\multicolumn{4}{|l|}{\small DC1--DC4 (Demotivators Main Categories/themes)} \\
\hline

\end{tabularx}
\end{table}

\begin{tcolorbox}[
colback=gray!12,
colframe=gray!45,
arc=3mm,
boxrule=0.8pt,
width=\columnwidth,
enhanced,
breakable,
title={\faKey\hspace{0.5em}Final Insight from Factor-Level Optimization},
fonttitle=\bfseries,
coltitle=black,
boxed title style={
    colback=gray!20,
    colframe=gray!45,
    arc=2mm
},
attach boxed title to top left={xshift=4mm,yshift=-2mm},
]

The factor-level analysis shows that the GA selected different
levels within each category under the fitted probability models and the defined
allocation score. Relatively higher levels were assigned to factors related to
conceptual learning, structured feedback, governance, and reliability. However,
these results represent exploratory familiarity--allocation scenarios. They do
not demonstrate that the selected factors are more educationally effective or
should receive greater institutional priority. Classroom and institutional
validation is required before applying them in practice.

\end{tcolorbox}
\section{Discussion}
\label{sec:Discussions}

The findings demonstrate both the educational benefits of LLMs
and the conditions required for their responsible use. Stakeholders recognize
that LLMs can support teaching and learning. However, they are also concerned
about academic integrity, over-reliance, reduced critical thinking, privacy,
unreliable outputs, and ethical issues. These concerns are consistent with
previous research on generative AI in higher education
\cite{cotton2024,Dwivedi2023}. Therefore, integrating LLMs is not simply a
matter of introducing a new educational tool. It requires balancing expected
learning benefits with potential risks, institutional safeguards, and the
specific educational context.

\begin{figure*}[t]
    \centering
    \includegraphics[width=\textwidth]{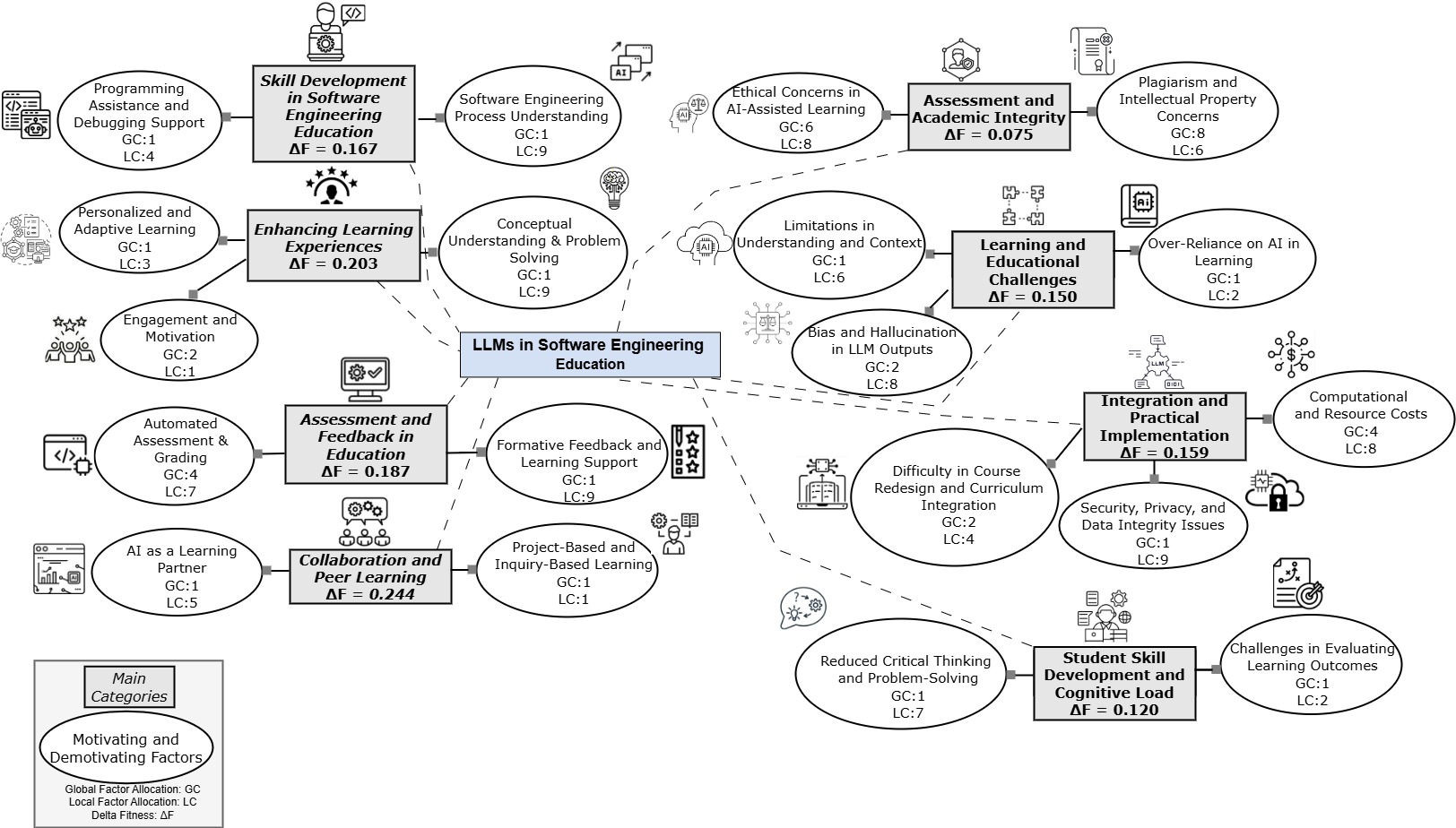}
    \caption{Exploratory Familiarity--Allocation Framework}
    \label{fig:prediction_taxonomy}
\end{figure*}

The GC, LC, and $\Delta$Fitness results given in Figure~\ref{fig:prediction_taxonomy} provide three related views of the analyzed
factors. GC compares all 19 motivating and demotivating factors under the
fitted probability models and the defined allocation score. In the
GA-generated global configuration, relatively higher levels were selected for
\textit{Plagiarism and Intellectual Property Concerns} and \textit{Ethical
Concerns in AI-Assisted Learning}. These results indicate that integrity- and
governance-related factors have a stronger influence on the identified
familiarity--allocation balance under the current model assumptions. They do
not demonstrate that these factors are universally more important or
educationally more effective. Therefore, the GC results should be interpreted
as model-based allocations rather than a universal ranking of factor
importance.

The Local Factor Allocation (LC) compares factors within their
individual categories. It identifies modeled allocation differences that may
not be clearly visible when all 19 factors are analyzed together. Within the
motivator categories, relatively higher levels were selected for
\textit{Conceptual Understanding and Problem Solving}, \textit{Software
Engineering Process Understanding}, and \textit{Formative Feedback and
Learning Support}. Within the demotivator categories, factors related to
reliability, privacy, governance, and critical thinking received relatively
higher levels. A factor with a lower allocation in GC may therefore still
receive a higher allocation within its own category through LC. These results
represent model-based familiarity--allocation patterns rather than validated
measures of practical or educational importance. The emphasis on cognitive
and process-oriented learning is consistent with the view that generative AI
can support learning rather than only automate tasks \cite{Kasneci2023}.

At the category level, $\Delta$Fitness measures the improvement
in fitness compared with the baseline configuration. The largest improvements
were observed for \textit{MC4\_Collaboration and Peer Learning} and
\textit{MC2\_Enhancing Learning Experiences}. The demotivating categories
generally produced smaller improvements under the fitted probability models
and the defined allocation score. These differences show how the fitness
function responds to alternative configurations within each category. They do
not directly measure classroom effectiveness, actual financial cost,
institutional effort, or causal effects on learning outcomes.

Together, the three analyses answer different questions. GC
shows the selected factor levels when all 19 factors are analyzed together. LC
shows the selected levels among factors within each category.
$\Delta$Fitness shows the model-based improvement relative to the baseline
configuration at the category level. Considering these three analytical levels
prevents the global configuration from hiding important within-category
patterns. The framework therefore converts stakeholder perceptions into
global, local, and baseline-relative familiarity--allocation scenarios.

The familiarity analysis provides an important limitation for
interpreting this framework. The modest classification performance indicates
that the 19 perception variables provide limited ability to distinguish
respondents with higher and lower levels of self-reported LLM familiarity.
This may suggest that similar benefits and concerns are reported across the
two familiarity groups, but it may also reflect limited predictive information
in the available data. Therefore, the fitted models should not be used to
predict an individual respondent's familiarity level. Their purpose in this
study is to provide estimated probability values for the optimization
framework. Consequently, the GA results inherit the uncertainty and
limitations of these probability models.

Overall, the framework extends the descriptive taxonomy by
providing an exploratory method for examining factor allocations under a
defined fitness function. It should be treated as a decision-support framework
rather than a validated implementation guideline. The results show how the
modeled familiarity--allocation balance changes across global, category, and
factor levels under the current assumptions. They do not demonstrate that
implementing the selected configurations will improve teaching, student
learning, or institutional LLM integration. Their practical and educational
value must be validated through classroom pilots, institutional case studies,
and replication using larger and more diverse samples.
\section{General Implications}
\label{sec:implications}

The study provides implications for researchers, instructors,
and institutions considering the integration of LLMs into software engineering
education. These implications are informed primarily by the stakeholder-evaluated taxonomy and the descriptive survey findings; the GA-generated configurations are considered only as exploratory inputs and are not treated as validated instructional priorities. They should be
interpreted as preliminary decision-support insights rather than validated
recommendations for educational or institutional practice.

Because the present survey was conducted to evaluate the
taxonomy and develop the exploratory framework, it did not include an
institutional implementation. To translate the framework into a testable
intervention, an institution could follow a five-step prospective pilot.
First, it should select a specific course or learning activity, define the
intended learning objectives, and record baseline values for the selected
outcomes. Second, instructors and relevant stakeholders should use the
taxonomy to identify locally relevant benefits and risks, treating the
GA-generated configurations as exploratory inputs alongside local expertise
rather than as prescribed priorities. Third, the institution should develop a
pilot protocol specifying permitted LLM-supported activities, instructor and
student training, disclosure requirements, output-verification procedures,
assessment arrangements, and safeguards for academic integrity, privacy,
intellectual property, and independent reasoning. Fourth, it should conduct a
limited-duration pilot using pre--post measurements and, where feasible, a
comparison group or comparable course offering. Fifth, the institution should
compare the results with predefined success criteria and decide whether to
revise, extend, or discontinue the intervention. Evaluation measures could
include course-assessment performance, performance on unaided
problem-solving tasks, code or test quality, students' ability to verify
LLM-generated outputs, academic-integrity incidents, instructor workload,
student and instructor feedback, and privacy or reliability incidents. This
pathway provides a concrete procedure for future institutional evaluation but
does not constitute practical validation within the present study.

\subsection{Academic and Theoretical Implications}

This study extends earlier descriptive research on AI in higher
education \cite{Zawacki2019} by combining three elements: a taxonomy of 19
motivating and demotivating factors, evaluation involving multiple stakeholder
groups, and an exploratory familiarity--allocation framework. This combination
provides a structured method for examining stakeholder perceptions and
comparing alternative factor configurations under the fitted probability models
and the defined allocation score.

The distinction between Global Factor Allocation (GC) and Local
Factor Allocation (LC) provides two complementary analytical views. GC examines
all 19 factors together, whereas LC compares factors within their individual
categories. Consequently, a factor receiving a lower allocation in GC may still
receive a higher allocation relative to other factors in its category through
LC. This multi-level perspective is consistent with technology-adoption
research, which views adoption as the result of multiple interacting factors
rather than a single determinant \cite{Venkatesh2003}.

The study also demonstrates how probabilistic modeling and
evolutionary optimization can be combined for exploratory decision support.
The fitted models are not intended to classify individual stakeholders or
establish causal relationships. Instead, they provide estimated probabilities
of high self-reported LLM familiarity that are used to compare alternative
factor configurations under the defined fitness function. Future research can
evaluate this approach using larger and more diverse samples, alternative
allocation assumptions, and data collected from classroom and institutional
settings.

\subsection{Practical Implications for Instructors}

The stakeholder-evaluated taxonomy provides instructors with a
structured checklist for planning LLM-supported teaching. It encourages
consideration of both expected educational benefits and potential challenges
before introducing LLMs into a software engineering course. Instructors can use
the taxonomy to examine whether planned activities support learning objectives
while also addressing academic integrity, reliability, privacy, and responsible
use.

The Local Factor Allocation (LC) results provide model-based
indications of factors that received relatively higher selected levels within
their respective pedagogical categories. These include \textit{Conceptual
Understanding and Problem Solving}, \textit{Software Engineering Process
Understanding}, and \textit{Formative Feedback and Learning Support}.
Instructors may consider these areas when designing initial LLM-supported
activities. However, the selected levels do not demonstrate that these factors
are more educationally effective than factors receiving lower levels. Their
value must be evaluated in actual teaching contexts.

The demotivating factors identify challenges that should also be
considered when planning LLM-supported activities. These include plagiarism and
intellectual property concerns, ethical concerns, unreliable or biased outputs,
over-reliance, reduced critical thinking, and security and privacy issues. The
taxonomy can support discussions about acceptable LLM use, disclosure of AI
assistance, verification of generated content, and students' responsibility for
their submitted work. The allocation results can help course teams compare
alternative areas of attention, but they should not be treated as a fixed
ranking of instructional priorities. 

Instructors may consider limited LLM integration when its use supports an explicit learning objective, students possess sufficient foundational knowledge to evaluate the generated output, and the instructor can monitor the activity. Suitable starting points include programming courses, once the relevant basic concepts have been introduced and the students have grasped them. Students should be able to identify, correct, and explain problems in the generated code. Other suitable contexts include software testing and quality-assurance courses in which students assess and improve generated test cases; software maintenance and code-review activities in which AI suggestions are compared with human reasoning; and project-based or capstone courses in which students document, verify, and reflect on their use of LLM-generated material. LLMs can also be used in Human–Computer Interaction (HCI) courses, provided that graphic design is not assessed as part of the module’s learning outcomes and the focus is on requirements gathering and on evaluating the usability and accessibility of an interface. In such courses, students can use LLMs to create digital prototypes quickly from low-fidelity paper prototypes. They then adapt these prototypes by applying the usability and accessibility principles they have learnt, refine the LLM output, and evaluate the resulting interfaces. In all of these contexts, assessment should emphasize the student’s explanation, verification, and modification of the output rather than the generated output alone. To this end, instructors need to define in detail how students should document the process. This documentation should include the prompts used, the outputs produced by the LLM, the errors made by the LLM, and the modifications made to the generated output, together with the criteria and reasons that make each modification necessary. 

LLM use should be restricted when the objective is to assess unaided competence, particularly in high-stakes examinations, when students are still developing essential foundational skills through direct practice, or when an activity involves confidential, personal, proprietary, or security-sensitive information. Before introducing LLMs, instructors should define permitted and prohibited uses, disclosure and citation requirements, output-verification responsibilities, privacy and intellectual-property safeguards, and procedures for maintaining independent reasoning and equitable access. These considerations are preliminary guidance derived from the stakeholder-evaluated taxonomy and survey findings rather than validated instructional recommendations.

\subsection{Institutional and Curriculum-Level Implications} \label{Sec:Intitutional Implications}

At the institutional level, the Global Factor Allocation (GC)
results assigned relatively higher levels to \textit{Plagiarism and Intellectual
Property Concerns} and \textit{Ethical Concerns in AI-Assisted Learning} under
the fitted probability models and the defined allocation score. These findings
may inform discussions about academic integrity policies, responsible AI
guidance, staff training, and support for LLM use. Similar governance concerns
have been identified in previous research on generative AI in higher education
\cite{cotton2024}. However, the GC results do not demonstrate that prioritizing
these factors will necessarily improve teaching, student learning, or
institutional LLM integration.

The taxonomy and category-level results can also support
decisions about the scope of initial institutional or curriculum-based pilots.
Rather than introducing LLMs widely from the beginning, institutions may test a
limited set of LLM-supported activities in one or more courses. Such pilots
could evaluate student learning, instructor and student experiences, academic
integrity, staff effort, training requirements, technical resources, and
unintended consequences before wider adoption.

These implications should be treated as preliminary
decision-support guidance. No institution or course in the present study
implemented the GA-generated configurations, and their effects on teaching,
curriculum design, institutional practices, and student learning were not
measured. Classroom pilots, institutional case studies, and replications with
larger and more diverse samples are therefore required to determine whether the
model-derived allocations are useful and effective in practice.
\section{Threats to Validity}
\label{sec:Threats}

Various factors may affect the validity of the study findings.
Following the empirical software engineering guidelines proposed by Wohlin
et al.~\cite{wohlin2012experimentation}, we discuss these threats in terms of
internal validity, external validity, construct validity, and conclusion
validity.

\subsection{Internal Validity}

Internal validity concerns whether methodological choices or
uncontrolled influences may have affected the observed results. This study
relies on self-reported survey data; therefore, participants' ratings may
reflect subjective perceptions, previous experience with LLMs, or their
institutional and disciplinary contexts. Response biases, including social
desirability and individual differences in interpreting the questions, may
also have influenced the reported ratings.

To reduce these threats, the survey instrument was grounded in
the literature-derived taxonomy developed in our earlier work
\cite{khan2025integrating}. It was refined through internal review, structured
piloting, and external expert evaluation to improve clarity, strengthen the
alignment between the identified factors and survey items, and reduce
interpretational ambiguity. The same survey structure and 9-point response
scale were presented consistently to all participants. Nevertheless,
self-reporting and contextual influences cannot be completely eliminated.

\subsection{External Validity}

External validity concerns the extent to which the findings can
be generalized beyond the studied sample. The study focuses on LLM integration
in software engineering education within higher education institutions.
Consequently, the findings may not directly apply to other educational
disciplines, professional software development environments, vocational
education, or institutions with substantially different technological and
organizational conditions. Differences in institutional size, AI maturity,
governance structures, teaching practices, and available resources may also
affect the applicability of the identified factors and GA-generated
configurations.

The participants represented multiple academic roles and
countries, providing perspectives from different institutional and national
contexts. This diversity reduces dependence on a single institutional
viewpoint but does not make the sample statistically representative of the
global higher education population. The non-probability
recruitment approach and voluntary participation may have introduced coverage
and self-selection bias, as individuals with greater interest in or experience
with LLMs may have been more likely to participate. Replication across
additional universities, countries, disciplines, and stakeholder groups is
therefore required. Longitudinal studies would also be valuable because
perceptions and institutional practices may change as LLM technologies and
related policies continue to evolve.

\subsection{Construct Validity}

Construct validity concerns whether the survey instrument
adequately represents the theoretical constructs being examined. The 19
motivating and demotivating factors were derived from a previous systematic
literature review and organized into a structured taxonomy. However, complex
constructs such as \textit{Conceptual Understanding and Problem Solving} and
\textit{Ethical Concerns in AI-Assisted Learning} may not be fully captured by
single survey items. Participants may also have interpreted some factors
differently depending on their academic roles, disciplines, experiences, or
institutional contexts.

To mitigate these threats, the survey items were explicitly
mapped to the taxonomy categories and sub-themes and refined through internal
review, piloting, and external expert feedback. However, encoding the 9-point
Likert-scale responses as numerical factor levels assumes that differences
between adjacent values can be treated consistently within the predictive and
optimization models. This assumption may not fully reflect the ordinal nature
of the responses.

The outcome variable represents high versus lower
self-reported LLM familiarity. It reflects participants' perceived knowledge
of and exposure to LLMs; it does not measure their competence in integrating
LLMs into curricula, institutional readiness, pedagogical effectiveness, or
student learning outcomes. Future research should use validated
multidimensional measures and combine surveys with interviews, observations,
or case studies to triangulate the interpretation of these constructs.

\subsection{Conclusion Validity}

Conclusion validity concerns the credibility of the statistical, predictive, and optimization-based inferences. The analysis included 19 predictors and 126 responses, with an imbalanced outcome distribution of 90 higher-familiarity and 36 lower-familiarity respondents. This sample may be insufficient for identifying subtle or stable differences between the two groups. Both Naive Bayes and Logistic Regression achieved an accuracy of 0.461, below the majority-class baseline. Therefore, the factors did not reliably distinguish between respondents with higher and lower levels of self-reported LLM familiarity, and the probability estimates should not be interpreted as reliable individual-level predictions. The GA consequently inherits the weak discriminatory performance and uncertainty of the underlying probability models. Its outputs demonstrate the behavior of the specified objective under the current assumptions rather than establish that the probability estimates are sufficiently reliable for prescriptive optimization.
The binary familiarity outcome also involves information
loss, and the study-specific cutoff affects the class distribution and the
models used by the GA. Alternative thresholds could therefore produce
different model estimates and GA configurations.

The models also rely on assumptions that may not fully hold in
the present data. Naive Bayes assumes conditional independence among
predictors, whereas Logistic Regression assumes a linear relationship between
the predictors and the log-odds of the outcome. Using two models and averaging
their estimated probabilities provides two complementary model perspectives,
but it does not remove their assumptions or compensate for their modest
predictive performance. Although a stratified 80/20 train--test split preserved
the outcome distribution, results may vary across alternative data partitions
or validation procedures. Some synthetic configurations may fall outside densely
represented regions of the observed data, where the estimated probabilities,
particularly those produced by Naive Bayes, may be unstable. Averaging the
Naive Bayes and Logistic Regression estimates does not eliminate this risk.
Consequently, the GA results should be interpreted as model-dependent
exploratory scenarios rather than stable optimization outcomes. Future
robustness studies should compare model-specific GA results and examine search
constraints that restrict candidate configurations to empirically supported
regions.

The GA is stochastic and may produce near-optimal rather than
globally optimal configurations. Its outputs depend on the estimated
probabilities, fitness-function design, allocation score, parameter settings,
and baseline configuration. Different outcome definitions, model estimates,
factor weightings, allocation assumptions, or GA parameters could therefore
produce different configurations. Reporting the modeling procedures and baseline comparisons improves transparency and reproducibility but does not establish the practical effectiveness of the selected configurations.

The allocation score is a normalized relative cost--effort proxy
and does not represent actual institutional expenditure. The allocation score also assumes that all factor levels contribute equally and linearly to the penalty. This simplified formulation does not capture factor-specific or nonlinear differences in institutional resource requirements, and the resulting configurations may change under alternative allocation assumptions. Consequently, GC and LC represent exploratory
familiarity--allocation scenarios rather than institutional budgets or
validated implementation priorities. Similarly, $\Delta$Fitness represents
improvement in the defined model objective relative to the baseline; it does
not measure improvement in teaching quality, student learning, institutional
readiness, or cost efficiency.

Finally, the cross-sectional design supports the identification
of associations but not causal conclusions. The study does not demonstrate
that changing a factor level will increase LLM familiarity or improve LLM
integration. Future studies should use larger independent samples,
multidimensional measures of LLM familiarity and integration readiness,
repeated or nested cross-validation, sensitivity analyses for alternative
allocation assumptions, and classroom or institutional evaluations of the
GA-generated configurations.

\section{Conclusions}
\label{sec:conclusion}

This study examined motivating and demotivating factors associated with LLM
use in software engineering education through an empirically grounded and
exploratory familiarity--allocation approach. Based on 126 valid responses
from multiple stakeholder groups, we evaluated nine motivating and ten
demotivating factors and examined their relationship with high self-reported
LLM familiarity. Naive Bayes and Logistic Regression provided estimated
probabilities that were incorporated into a GA-based framework using a defined
allocation score. The framework supported Global Factor Allocation (GC), Local
Factor Allocation (LC), and category-level $\Delta$Fitness analyses.

The descriptive results showed that \textit{Programming Assistance and
Debugging Support}, \textit{Personalized and Adaptive Learning}, and
\textit{AI as a Learning Partner} were the most highly rated motivators.
\textit{Plagiarism and Intellectual Property Concerns}, \textit{Over-Reliance
on AI in Learning}, and \textit{Reduced Critical Thinking and Problem-Solving}
were the most highly rated demotivators. These findings indicate that
stakeholders recognize the educational potential of LLMs while remaining
concerned about their implications for academic integrity, independent
learning, and critical thinking.

However, both probabilistic models achieved an accuracy of 0.461, below the
majority-class baseline. Therefore, the 19 factor ratings did not reliably
distinguish respondents with high self-reported LLM familiarity from those
with moderate or lower familiarity. The resulting probability estimates
should consequently be interpreted as exploratory relative scoring functions
rather than reliable individual-level predictions. The high-familiarity
cutoff is also a transparent, study-specific operationalization rather than a
universal or externally validated threshold.

Under the fitted probability models and the defined allocation score, the GA
produced different configurations across the three analytical levels. GC
assigned relatively higher levels to integrity- and governance-related
factors, particularly \textit{Plagiarism and Intellectual Property Concerns}
and \textit{Ethical Concerns in AI-Assisted Learning}. LC revealed
within-category patterns involving conceptual learning, software engineering
process understanding, structured feedback, reliability, privacy, governance,
and critical thinking. Category-level $\Delta$Fitness showed the improvement
in the defined model objective relative to the baseline configuration. These
results answer the research question at an exploratory level by demonstrating
how alternative factor configurations can be compared under a
familiarity--allocation objective. They do not establish causal relationships,
actual institutional expenditure, pedagogical effectiveness, competence in
curriculum integration, institutional readiness, or successful LLM adoption.

The main contribution of this study is a structured framework connecting a
literature-derived taxonomy of 19 motivating and demotivating factors,
multi-stakeholder perceptions, probabilistic modeling, and evolutionary
optimization. The resulting configurations provide preliminary
decision-support insights rather than validated implementation priorities.
Future research should evaluate the framework using larger and more diverse
independent samples,examine the robustness of the familiarity
operationalization through systematic threshold-sensitivity analyses and
ordinal familiarity models, and develop validated multidimensional measures
of LLM familiarity and integration readiness.
Future research should also examine the sensitivity of the
GA results under expert-informed or institution-specific weighted and
nonlinear allocation formulations that reflect differences in staff time,
training, licensing, infrastructure, policy development, technical support,
and other institutional resources.
Finally, prospective classroom pilots and institutional case
studies should implement the operationalization pathway presented in
Section~\ref{sec:implications} and evaluate predefined learning, integrity,
workload, and reliability outcomes to determine whether the framework can
support teaching, student learning, and responsible LLM integration in
practice.
\clearpage
\noindent\textbf{Credit Authorship Contribution Statement:}

\textbf{Maryam Khan}: Conceptualization, Methodology, Investigation, Data Curation, Formal Analysis, Software, Validation, Visualization, Writing -- Original Draft. \textbf{Maryam Khan} led the study, developed the survey instrument, conducted data collection, performed data preprocessing and modeling, implemented the optimization framework, and prepared the initial manuscript draft. \textbf{Jussi Kasurinen}: Validation, Supervision, Review \& Editing. \textbf{Jussi Kasurinen} reviewed and validated the survey instrument, contributed to survey distribution and data collection support, supervised the research activities, and provided feedback and revisions on the manuscript. \textbf{Muhammad Azeem Akbar}: Writing -- Review \& Editing. \textbf{Muhammad Azeem Akbar} reviewed the manuscript and provided feedback on its content and presentation. \textbf{Estefanía Martín-Barroso}: Review \& Editing. \textbf{Estefanía Martín-Barroso} reviewed the manuscript and provided feedback on the reviewers' comments and response files.

\noindent\textbf{Declaration of competing interest:} 
The authors declare that they have no known competing financial interests or personal relationships that could have appeared to influence the work reported in this paper.

 \noindent\textbf{Acknowledgments:} 

The authors would like to sincerely thank all survey respondents who voluntarily participated in this study and contributed their time and insights. We also gratefully acknowledge the external expert who reviewed the survey instrument and provided constructive feedback during the piloting phase, which significantly improved the clarity and structure of the questionnaire. The authors further acknowledge the use of AI assisted tools, including ChatGPT and Grammarly, for language polishing and improving the clarity of the manuscript. These tools were used solely to enhance writing quality and readability. The authors take full responsibility for the study design, data collection, data analysis, modeling, results, and conclusions presented in this paper.

\bibliographystyle{unsrtnat}
\bibliography{References}

\end{document}